\documentclass[a4paper, aps, prd, amsfonts, floatfix, amssymb, amsmath, reprint, showkeys, nofootinbib, oneside, superscriptaddress, preprintnumbers]{revtex4-1}
\usepackage[english]{babel}
\usepackage[utf8]{inputenc}
\usepackage{slashed}
\usepackage{lipsum} 
\usepackage[force]{feynmp-auto}
\usepackage{graphicx}
\usepackage{dcolumn}
\usepackage{bm}
\usepackage{color}
\usepackage{multirow}
\usepackage{textgreek}
\usepackage{tabularx}
\usepackage{slashed}
\newcommand{\pq}{\bar{P}\cdot\bar{q}}
\usepackage{multirow, makecell}
\newcolumntype{C}{>{\centering\arraybackslash}X}
\usepackage[colorinlistoftodos, color=green!40, prependcaption]{todonotes}
\usepackage{amsthm}
\usepackage{mathtools}
\usepackage{physics}
\usepackage{xcolor}
\usepackage{graphicx}
\usepackage[left=23mm,right=13mm,top=35mm,columnsep=15pt]{geometry} 
\usepackage{adjustbox}
\usepackage{placeins}
\usepackage[T1]{fontenc}
\usepackage{lipsum}
\usepackage{csquotes}
\usepackage[normalem]{ulem}

\usepackage[pdftex, pdftitle={Article}, pdfauthor={Author}]{hyperref} 
\DeclareUnicodeCharacter{2009}{\,} 

\renewcommand{\>}{\rangle}
\newcommand{\<}{\langle}
\newcommand{\bomega}{\bar{\omega}}

\begin{document}
\title{Reconstructing generalised parton distributions from the lattice off-forward Compton amplitude}

\author{A.~Hannaford-Gunn}
    \affiliation{CSSM, Department of Physics, The University of Adelaide, Adelaide SA 5005, Australia}
\author{K.~U.~Can}
    \affiliation{CSSM, Department of Physics, The University of Adelaide, Adelaide SA 5005, Australia}
    \author{J.~A.~Crawford}
    \affiliation{CSSM, Department of Physics, The University of Adelaide, Adelaide SA 5005, Australia}
    \author{R.~Horsley}
    \affiliation{School of Physics and Astronomy, University of Edinburgh, Edinburgh EH9 3JZ, UK}
\author{P.~E.~L.~Rakow}
    \affiliation{Theoretical Physics Division, Department of Mathematical Sciences, University of Liverpool, Liverpool L69 3BX, UK}
    \author{G.~Schierholz}
    \affiliation{Deutsches Elektronen-Synchrotron DESY, Notkestr.~85, 22607 Hamburg, Germany}
      \author{H.~St\"uben}
    \affiliation{Regionales Rechenzentrum, Universit\"at Hamburg, 20146 Hamburg, Germany}
    \author{R.~D.~Young}
    \affiliation{CSSM, Department of Physics, The University of Adelaide, Adelaide SA 5005, Australia}
    \author{J.~M.~Zanotti}
    \affiliation{CSSM, Department of Physics, The University of Adelaide, Adelaide SA 5005, Australia}

    \collaboration{CSSM/QCDSF/UKQCD Collaborations}
	\noaffiliation

\date{} 

\begin{abstract}
We present a determination of the structure functions of the off-forward Compton amplitude $\mathcal{H}_1$ and $\mathcal{E}_1$ from the Feynman-Hellmann method in lattice QCD. At leading twist, these structure functions give access to the generalised parton distributions (GPDs) $H$ and $E$, respectively. This calculation is performed for an unphysical pion mass of $m_{\pi}=412\;\text{MeV}$ and four values of the soft momentum transfer, $t\approx 0, -0.3, -0.6, -1.1\;\text{GeV}^2$, all at a hard momentum scale of $\bar{Q}^2\approx 5\;\text{GeV}^2$. Using these results, we test various methods to determine properties of the real-time scattering amplitudes and GPDs: (1) we fit their Mellin moments, and (2) we use a simple GPD ansatz to reconstruct the entire distribution. Our final results show promising agreement with phenomenology and other lattice results, and highlight specific systematics in need of control. 
\end{abstract}

\preprint{ADP-24-08/T1247}
\preprint{DESY-24-065}
\preprint{Liverpool LTH 1370}
\maketitle

\section{Introduction}
\label{sec:intro}

Generalised parton distributions (GPDs) are among the most important hadron structure observables, giving access to the spatial probability distributions of quarks \cite{burkardt}, as well as the spin \cite{jiog}, force and pressure distributions within a hadron \cite{mechanprops0, dtermexp}.

While GPDs can in principle be determined from hard exclusive scattering processes, in practice such experimental determinations face certain difficulties \cite{gpdphenomreview, gpdphenomreview2, deconvolution}. In this context, first principles calculations of GPDs from lattice QCD can be extremely useful in both guiding and comparing to experiment. There have already been a handful of studies exploring the ability of lattice results to aid experimental GPD fits \cite{neuralnetwork_latticemoms, ji_lattice_phenom_fit, ji_lattice_phenom_fit_xi, pseudo_plus_expt, stringmodel_v_lqcd}. 

While lattice calculations cannot directly determine GPDs, they can be used to reconstruct them. Historically, this has been limited to determinations of the first three Mellin moments of GPDs, calculated from local twist-two operators \cite{gpdlatt1, gpdlatt2,gpdlatt3, gpdlatt4,gpdlatt6,gpdlatt7,gpdlatt8,gpdlatt9}. Recent advances since 2015 have opened new avenues for reconstructing GPDs from non-local operators, notably the quasi- \cite{jiquasi} and pseudo-distribution \cite{radpseudo} approaches, which have yielded promising initial determinations of GPDs \cite{pionquasi, nucleonquasi1, nucleonquasi2, pseudogpds, quasiGPDpion2}.

In this paper, we apply the Feynman-Hellmann method to determine the off-forward Compton amplitude (OFCA) and thereby access GPDs. A major distinction between this method and those mentioned above is that we calculate a lattice version of the scattering amplitude from which GPDs are determined experimentally. As such, we have the potential to calculate phenomenologically interesting properties that are inaccessible to other methods. 

So far, the \emph{forward} Compton amplitude has been the focus of the Feynman-Hellmann calculations \cite{fwdletter, fwdpaper}, including studies of the scaling behaviour and higher-twist structures \cite{comptonproceedings, fwdpaperscaling} and the subtraction function \cite{interlacingsubtraction}. Similarly, a calculation of the {off-forward} Compton amplitude could determine the $\bar{Q}^2$ scaling of this amplitude, of which there have only been limited experimental studies \cite{dvcsscalingtest1, dvcsscalingtest2, dvcshighertwist1, dvcshighertwist2}. For the OFCA the $\bar{Q}^2$ scaling behaviour could be extremely useful in isolating the leading-twist contribution, as most deeply virtual Compton scatting experiments have a relatively modest hard scale of $Q^2\approx 1-12\;\text{GeV}^2$ and contain additional $|t|/Q^2$ corrections \cite{dvcshttheory1, dvcshttheory2}. Moreover, this method can determine the off-forward subtraction function, which is a key input for experimental determinations of the proton pressure distribution \cite{dtermexp, pressuredistcomment}. 

In this work we focus on determining the subtracted structure functions of the OFCA for a single hard scale, $\bar{Q}^2\approx 5\;\text{GeV}^2$. We build upon our previous paper on the OFCA \cite{hannafordgunn2021generalised}, where we developed the formalism for this calculation and presented exploratory numerical results. Here, we present major improvements on this numerical calculation: we separately determine the $\mathcal{H}_1$ and $\mathcal{E}_1$ structure functions, which can be uniquely related to twist-two GPDs, over a wider range of kinematics. These improvements allow us to focus on modelling and determining GPD properties: (1) we determine the Mellin moments of $\mathcal{H}_1$ and $\mathcal{E}_1$ with a largely model-independent fit, and (2) fit our Compton amplitude results using a phenomenologically motivated GPD ansatz.

\section{Background} \label{sec:bckgrnd}

Our purpose in this paper is to determine the off-forward Compton amplitude (OFCA) and its structure functions from lattice QCD. The notation of this paper follows that of Ref.~\cite{hannafordgunn2021generalised}, in which can be found derivations and full expressions for analytic results used here.

\begin{figure}[t!]
\centering
\includegraphics[width=0.64\linewidth]{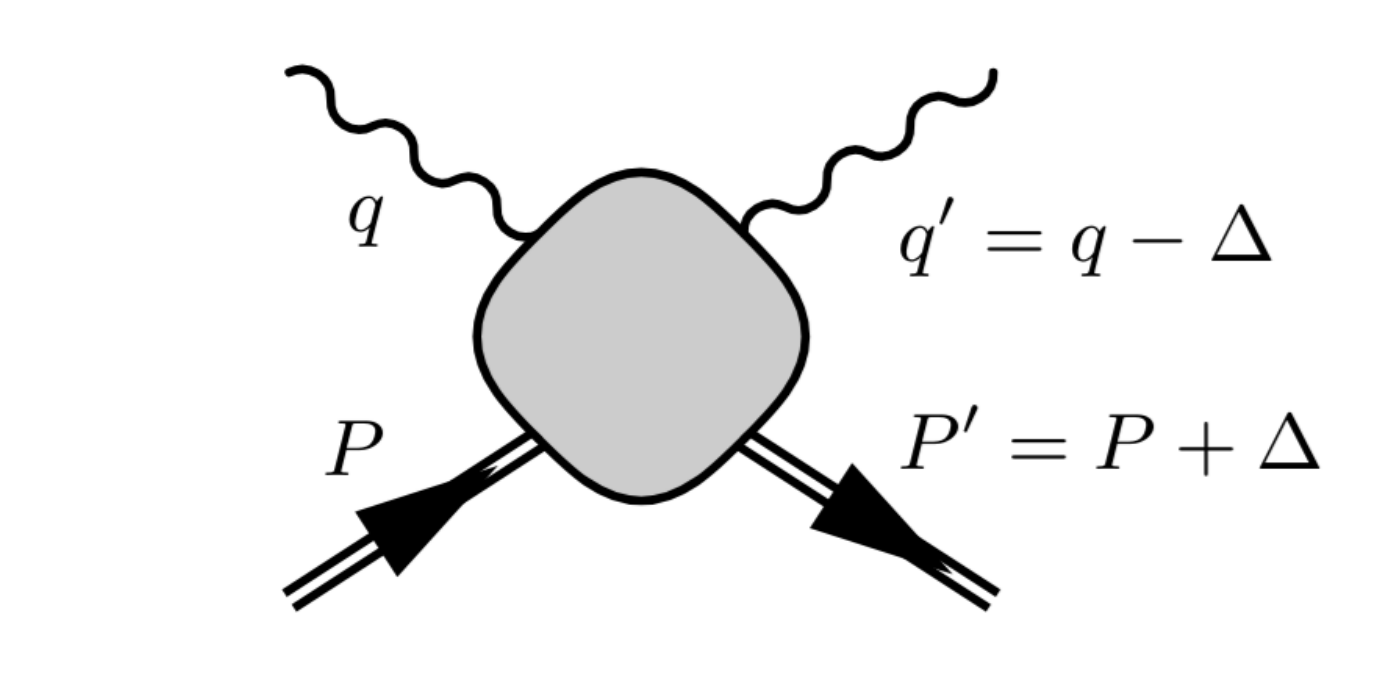}
    \caption{Diagram of off-forward $\gamma^{*} (q)N(P)  \to \gamma^{*}(q')N(P') $ scattering.}
    \label{offfwdpic}
    \end{figure}

The OFCA is defined as 
\begin{align}
       T_{\mu\nu}\equiv i\int d^4ze^{\frac{i}{2}(q+q')\cdot z} \<P'|T\{j_{\mu}(z/2)j_{\nu}(-z/2)\}|P\>,
       \label{ofcadef}
   \end{align} 
where $j_{\mu}$ is the electromagnetic current operator. See Fig.~\ref{offfwdpic} for the diagram of this process. 

In terms of kinematics, we use the basis of momentum vectors
\begin{align}
         \bar{P}= \frac{1}{2}(P+P'), \quad \bar{q} = \frac{1}{2}(q+q'), \quad \Delta = P' - P,
         \label{vectordef}
\end{align}
which gives us two scaling variables and two momentum scales:
\begin{align}
    \bomega = \frac{2\bar{P}\cdot\bar{q}}{-\bar{q}^2}, \quad \xi=\frac{\Delta\cdot \bar{q}}{2\bar{P}\cdot\bar{q}}, \quad \bar{Q}^2 = -\bar{q}^2, \quad t = \Delta^2.
    \label{scalardef}
\end{align}
For this work, we focus on zero-skewness kinematics, $\xi=0$, which is enforced by choosing $\bar{q}\cdot \Delta=0$.

The general OFCA is parametrised by 18 linearly independent {off-forward structure functions} \cite{perrottet, tarrach}\begin{align}
{T}_{\mu\nu} & = -\frac{1}{2\pq}\Big ( h\cdot \bar{q} \mathcal{H}_1+ e\cdot \bar{q} \mathcal{E}_1  \Big) \left(g_{\mu\nu}-\frac{q'_{\mu}q_{\nu}}{q\cdot q'} \right) + \dots
    \label{compamp_td}
\end{align}
where we have defined the Dirac bilinears
\begin{equation}
h_{\mu}=\bar{u}(P')\gamma_{\mu}u(P) \text{,  and  } e_{\mu}=\bar{u}(P')\frac{i \sigma_{\mu\kappa}\Delta^{\kappa}}{2m_N}u(P).
    \label{hebilineardef}
\end{equation}
See Ref.~\cite{hannafordgunn2021generalised} for a full parametrisation of the Compton amplitude; the tensor decomposition with all zero-skewness structures is given in Eq.~\eqref{expltensordecomp2}. For this work, we are interested in the helicity-conserving $\mathcal{H}_1$ and helicity-flipping $\mathcal{E}_1$ structure functions, which respectively give access to the $H$ and $E$ GPDs at leading-twist. Our structure functions $\mathcal{H}_1$ and $\mathcal{E}_1$ are comparable to the $\mathcal{H}$ and $\mathcal{E}$ Compton form factors used in experimental analyses \cite{belitskymullerkirchner, dv2quasireal, CFFsParam}.

The Euclidean Compton amplitude we calculate in lattice QCD can only be related to the Minkowski amplitude if we use the kinematics $|\bomega|< 1$ \cite{fwdpaper}. However, the real-time scattering amplitude at zero-skewness has the kinematics $|\bomega|> 1$.

As such, we use a dispersion relation to connect our lattice results ${\mathcal{H}}_1(\bomega, t, \bar{Q}^2)$ to the real-time structure function, $\mathcal{H}_1(x, t, \bar{Q}^2)$:
\begin{align}
  \begin{split}
\overline{\mathcal{H}}_1(\bomega, t, \bar{Q}^2)& =  \frac{2\bomega^2}{\pi}\int_{0} ^1 dx \frac{x \text{Im}\mathcal{H}_1(x, t, \bar{Q}^2)}{1-x^2\bomega^2 },
      \label{subDR}
  \end{split}
\end{align}
where $\overline{\mathcal{H}}_1(\bomega,t,\bar{Q}^2) = {\mathcal{H}}_1(\bomega,t,\bar{Q}^2) - {\mathcal{H}}_1(0,t,\bar{Q}^2)$, the subtracted structure function. An analogous result can be derived for the $\mathcal{E}_1$ structure function. 

At $t=0$, the forward limit, $\mathcal{H}_1$ reduces to the forward Compton structure function, $\mathcal{F}_1$, and one may use the optical theorem to write Eq.~\eqref{subDR} as
\begin{align}
    \overline{\mathcal{F}}_1(\omega, {Q}^2) ={4\omega^2}\int_0^{1}dx\frac{xF_1(x, Q^2)}{1-x^2\omega^2},
\end{align}
where $F_1$ is the deep inelastic scattering structure function.

In the leading-twist approximation ($\bar{Q}^2\gg\Lambda_{\text{QCD}}^2$), Eq.~\eqref{subDR} becomes \cite{hannafordgunn2021generalised}
\begin{align}
  \begin{split}
        \overline{\mathcal{H}}_1(\bomega, t)& =  2\bomega^2\int_{-1} ^1 dx \frac{x {H}(x, t)}{1-x^2\bomega^2},
      \label{DR_LT}
  \end{split}
\end{align}
where $H(x,t)$ is the twist-two helicity-conserving GPD. An analogous result exists for the replacements $\mathcal{H}_1\to \mathcal{E}_1$ and ${H}\to {E}$.

From Eqs.~\eqref{subDR} and \eqref{DR_LT} we see that it is in principle possible to reconstruct the real-time scattering amplitude and the GPDs from our Euclidean Compton amplitude. However, such a reconstruction is hampered by the fact that Eqs.~\eqref{subDR} and \eqref{DR_LT} have the form of a Fredholm integral equation of the first kind, which is an ill-conditioned inverse problem known to have numerically unstable solutions \cite{montecarlofit}. 

As such, in this work we employ two strategies to access the structure functions---similar techniques have been explored for the quasi- and pseudo-distribution methods \cite{pseudogpds, pseudodist_inverse, pseudo_pion}. First, we expand Eq.~\eqref{subDR} about $\bomega=0$, which gives us a power series in its Mellin moments. By varying $\bomega$, we can then determine a finite set of the moments. Such an extraction of structure function moments has been performed for the forward Compton amplitude \cite{fwdletter, fwdpaper, fwdpaperscaling} and in an exploratory calculation of the OFCA \cite{hannafordgunn2021generalised}. 

While such a fit on the level of moments gives us a largely model-independent determination, it is difficult to reconstruct the full structure function/GPD from a limited number of moments. As such, for our second fit method we use a phenomenologically-motivated GPD ansatz. In this work, we take
\begin{align}
    H(x,t) \propto x^{-\alpha_0-\alpha' t}(1-x)^{\beta}.
\end{align}
Similar parametrisation have been widely used in both the phenomenological studies of GPDs \cite{reggeparam1, gpdmodel2004, diehlmodel, FFfromGPD, gpdmodel2006, valenceregge1, reggeparam_dvcs, valenceregge2, globalfitgonzalez, valenceregge3, valenceregge4} and fits to lattice results from other methods \cite{ji_lattice_phenom_fit, ji_lattice_phenom_fit_xi}. Our implementation of this model-dependent fit follows previous work trialled in the forward case \cite{montecarlofit, utku_xfit}.

For both fitting approaches, we emphasise that the purpose is not to perform a perturbative matching and extract  a leading-twist GPD. Instead, our long-term goal is a determination of the moments and real-time structure function \emph{including their power corrections and higher-twist effects}. Such corrections provide us with useful information about the scale dependence of the amplitude, which has not been studied experimentally in great detail for off-forward Compton scattering. However, for the present work we perform calculations for a single $\bar{Q}^2$ value to focus on determining the structure functions and their moments.

	\begin{table*}[t!]
				\centering
				\caption{ \label{tab:gauge_details} Details of the gauge ensembles used in this work.}
				\setlength{\extrarowheight}{2pt}
	    	\begin{tabularx}{\textwidth}{CCCCCCCCCCC}
					\hline\hline
					$N_f$ & $c_{SW}$ & $\beta$ & $\kappa_l, \kappa_s$& $N_L^3 \times N_T$ & $a$ & $m_\pi$   & $Z_V$ & $N_\text{cfg}$\\
					&&&&& [fm] &[MeV]&& \\
					\hline
					$2+1$ & 2.48 & 5.65 & 0.122005  & $48^3\times96$ & 0.068 & $412$  & 0.871 & 537 \\
					\hline\hline
				\end{tabularx}
			\end{table*}

\section{Determination of the off-forward Compton amplitude} \label{sec:OFCA}

To calculate the Compton amplitude in lattice QCD, we use the Feynman-Hellmann (FH) method, which has proven to be a powerful tool to compute matrix elements with one and two operator insertions, and a useful alternative to the direct computation of three- and four-point functions.

This is implemented on the level of quark propagators, which are perturbed by two background fields:
\begin{align}
   S_{(\lambda_1, \lambda_2)}(x_n - x_m)= \big [M - \lambda_1\mathcal{O}_1- \lambda_2\mathcal{O}_2 \big]^{-1}_{n,m},
\end{align}
where $M$ is the Wilson fermion matrix and $\lambda_{1,2}$ are the FH couplings.

The perturbing operators are 
\begin{align}    \big[\mathcal{O}_{i}\big]_{n,m} =  \delta_{n,m} (e^{i\mathbf{q}_i\cdot\mathbf{z}_n}+e^{-i\mathbf{q}_i\cdot\mathbf{z}_n})\boldsymbol{\gamma}\cdot \mathbf{\hat{e}}_k,\; i=1,2,
\label{pert_mat}
\end{align}
where ${\mathbf{q}}_{1,2}$ are the inserted momentum, and $\hat{\mathbf{e}}_k$ picks the direction of the vector current.

From these perturbed two-point quark propagators, we then construct a perturbed nucleon correlator: \begin{equation}
     \begin{split}
        &  \mathcal{G}^{\Gamma}_{\boldsymbol{\lambda}} (\tau, \mathbf{p}')= \sum_{\mathbf{x}} e^{-i\mathbf{p}'\cdot\mathbf{x}}\Gamma_{\beta\alpha}\prescript{}{\boldsymbol{\lambda}}{\<}\Omega|\chi_{\alpha} (\mathbf{x}, \tau)\bar\chi_{\beta}(0)|\Omega\>_{\boldsymbol{\lambda}},
             \label{pertprop}
     \end{split}
\end{equation}
where $\bar\chi$ and $\chi$ are the nucleon creation and annihilation operators, $\mathbf{p}'$ is the sink momentum, $\tau$ is the Euclidean time, $\Gamma$ is a spin-parity projector, $\boldsymbol{\lambda}=(\lambda_1,\lambda_2)$ and $|\Omega\>_{\boldsymbol{\lambda}}$ the perturbed vacuum.

These perturbed nucleon propagators can be related to the OFCA by the Feynman-Hellmann like relation \cite{hannafordgunn2021generalised}
\begin{align}
\begin{split}
    &  R^{\Gamma}_{\lambda}(\mathbf{p}', \tau)
     \overset{\tau\gg a}{\simeq} \frac{\tau \lambda^2}{2E_N(\mathbf{p}')} \mathcal{R}^{\Gamma}_{\mu\mu},
    \label{FHneat}
\end{split}
\end{align}
where $R^{\Gamma}_{\lambda}$ is the following combination of perturbed and unperturbed nucleon correlators
\begin{equation}
    R^{\Gamma}_{\lambda} = \frac{\mathcal{G}^{\Gamma}_{(\lambda,\lambda)}+\mathcal{G}^{\Gamma}_{(-\lambda,-\lambda)}  -\mathcal{G}^{\Gamma}_{(\lambda,-\lambda)}-\mathcal{G}^{\Gamma}_{(-\lambda,\lambda)}}{4\mathcal{G}^{\Gamma_{\text{unpol}}}_{(0,0)}},
    \label{combocorr}
\end{equation}
at some sink momentum $\mathbf{p}'$ and sink time $\tau$, and $\lambda$ is the magnitude of the Feynman-Hellmann coupling. 

The quantity $ \mathcal{R}^{\Gamma}_{\mu\nu}$ is defined as
     \begin{align}
\mathcal{R}^{\Gamma}_{\mu\nu} = \frac{\sum _{s,s'} \text{tr}\big[\Gamma u(P',s')T_{\mu\nu}  \bar{u}(P,s)\big]}{\sum _{s}{\text{tr}[\Gamma_{\text{unpol}} u(P',s)\bar{u}(P',s)]}},
     \label{spindeplattR}
\end{align}
and $T_{\mu\nu}$ is the OFCA.

The spin-parity projectors we use are defined in Euclidean space as
\begin{equation}
    \Gamma_{\text{unpol}} = \frac{1}{2}(\mathbb{I} + \gamma_4), \quad  \Gamma_{\text{pol}-\hat{\mathbf{e}}} = -\frac{i}{2}(\mathbb{I} + \gamma_4)\hat{\mathbf{e}}\cdot \boldsymbol{\gamma}\gamma_5.
    \label{spinparproj}
\end{equation}

We insert the perturbed quark propagators either for the doubly-represented ($u$) quarks, or the singly-represented ($d$) quarks individually. This means that in practice we calculate 
\begin{align}
       T_{\mu\nu}^{ff}= i\int d^4ze^{i\bar{q}\cdot z} \<P'|T\{j^f_{\mu}(z/2)j^f_{\nu}(-z/2)\}|P\>,
       \label{ofca_flavours}
   \end{align} 
   where $j^f_{\mu}=Z_V \bar{\psi}_f \gamma_{\mu}\psi_f$, the local vector current with flavour $f=u$ or $d$, and renormalisation factor $Z_V$. For most analytic expressions we suppress flavour indices; however, we include them for our numerical results.

The kinematics of the Compton amplitude are then completely derived from our sink momentum, $\mathbf{p}'$, and our two inserted momenta, $\mathbf{q}_{1,2}$. For instance our momentum vectors, Eq.~\eqref{vectordef}, become
\begin{align}
    \bar{\mathbf{p}} = \frac{1}{2}(\mathbf{p}' + \mathbf{p}), \quad  \bar{\mathbf{q}} = \frac{1}{2}(\mathbf{q}_1+\mathbf{q}_2), \quad \mathbf{\Delta} = \mathbf{q}_1-\mathbf{q}_2, 
\end{align}
where $\mathbf{p}=\mathbf{p}'- \mathbf{\Delta}$ is the source momentum. From these, the momentum scalars in Eq.~\eqref{scalardef} become
\begin{align}
    \begin{split}
        \bomega= \frac{2 \bar{\mathbf{p}}\cdot  \bar{\mathbf{q}}}{   \bar{\mathbf{q}}^2}, \quad \bar{Q}^2  = \bar{\mathbf{q}}^2, \quad t  = -\mathbf{\Delta}^2.
    \label{q1q2_t_Q2}
    \end{split}
\end{align}
To ensure the skewness variable is zero, we keep $\mathbf{\Delta}\cdot\bar{\mathbf{q}}=0$, or equivalently $|{\mathbf{q}}_1| = |{\mathbf{q}}_2|$.

The key kinematic choice made in this work in contrast to Ref.~\cite{hannafordgunn2021generalised} is to keep $\mathbf{\hat{e}}_k \propto {\mathbf{q}}_1 - {\mathbf{q}}_2 \equiv \mathbf{\Delta} $, where $\mathbf{\hat{e}}_k$ picks the current direction in Eq.~\eqref{pert_mat}. Note that $\mathbf{\hat{e}}_k$ is always chosen to be a purely spatial vector. This means our vector current, $j_{\mu}$, is collinear with the soft momentum transfer, $\Delta_{\mu}$.

\begin{table}[t!]
    \centering
    \caption{Current insertion momenta, $\mathbf{q}_{1,2}$, and derived kinematics for the four sets of correlators.}
    \setlength{\extrarowheight}{5pt}
					{
			\begin{tabularx}{.485\textwidth}{C|C|C|C|C|C}
				\hline\hline
			 		$\frac{L}{2\pi}\mathbf{q}_1$,	$\frac{L}{2\pi}\mathbf{q}_2$ & $\frac{L}{2\pi}\mathbf{\Delta}$ & $\frac{L}{2\pi}\bar{\mathbf{q}}$ & $t$ \newline $[\text{GeV}^2]$ &  $\bar{Q}^2$ $[\text{GeV}^2]$ &$N_{\text{meas}}$ 	\\
				\hline
					$(5,3,0)$ & --- & --- & $0$ & $4.86$ & $1605$ \\ 
				\hline
				 $(4,3,3)$ $(3,4,3)$ & $(1,-1,0)$ & $(\frac{7}{2},\frac{7}{2},3)$ & $-0.29$ & $4.79$ & $1031$ \\ 
                           \hline
				$(5,3,1)$ $(5,3,-1)$ & $(0,0,2)$ & $(5,3,0)$ & $-0.57$ & $4.86$ & $1072$ \\ 
				\hline
				 $(4,2,4)$ $(2,4,4)$ & $(2,-2,0)$ & $(3,3,4)$ & $-1.14$ & $4.86$ & $1031$ \\ 
				\hline
				\hline
			\end{tabularx}
			}
    \label{tab:q_kinematics}
\end{table}

As shown in Appendix \ref{sec:heisolation}, the choice $\mathbf{\hat{e}}_k \propto  \mathbf{\Delta} $ eliminates all structure functions except $\mathcal{H}_1$ and $\mathcal{E}_1$ from the $\mu=\nu=k$ OFCA:
\begin{align}
     \begin{split}
        {T}_{kk} & = -\frac{1}{2\pq}\Big ( h\cdot \bar{q}\mathcal{H}_1
      + e\cdot \bar{q}\mathcal{E}_1 \Big),
        \label{finalofcatensor}
        \end{split}
\end{align}
where $h_{\mu}$ and $e_{\mu}$ are the bilinears defined in Eq.~\eqref{hebilineardef}; recall that $k$ is always a spatial direction (i.e.~it has no temporal component). This is a major improvement on our previous work \cite{hannafordgunn2021generalised}, where only a linear combination of the $\mathcal{H}_{1,2,3}$ and $\mathcal{E}_{1,2}$ was accessible, and we were required to make leading-twist approximations to isolate these.

Inserting Eq.~\eqref{finalofcatensor} into the ratio of spin-parity traces given in Eq.~\eqref{spindeplattR}, we obtain
\begin{equation}
    \mathcal{R}^{\Gamma}_{kk} = N^{\mathcal{H}}_{\Gamma}\mathcal{H}_1 + N^{\mathcal{E}}_{\Gamma}\mathcal{E}_1,
\end{equation}
where the $N$ factors are from the bilinear coefficients of Eq.~\eqref{finalofcatensor} inserted into the traces of Eq.~\eqref{spindeplattR}.

Therefore, by using the two spin-parity projectors of Eq.~\eqref{spinparproj}, we determine $\mathcal{H}_1$ and $\mathcal{E}_1$ from the matrix equation:
\begin{equation}
    \left( \begin{array}{c} \mathcal{R}^{\text{unpol}}_{kk} \\  \mathcal{R}^{\text{pol}}_{kk}  \end{array} \right) =   \left( \begin{array}{cc} N^{\mathcal{H}}_{\text{unpol}} & N^{\mathcal{E}}_{\text{unpol}} \\ N^{\mathcal{H}}_{\text{pol}} & N^{\mathcal{E}}_{\text{pol}} \end{array} \right)\left( \begin{array}{c} \mathcal{H}_1  \\ \mathcal{E}_1  \end{array} \right).
    \label{HEsepmatrix}
\end{equation}
The matrix of $N$ factors has a determinant $\sim1$ for all our kinematics, making the inversion practical in all cases.

\begin{figure}[t!]
    \centering
    \includegraphics[width=\linewidth]{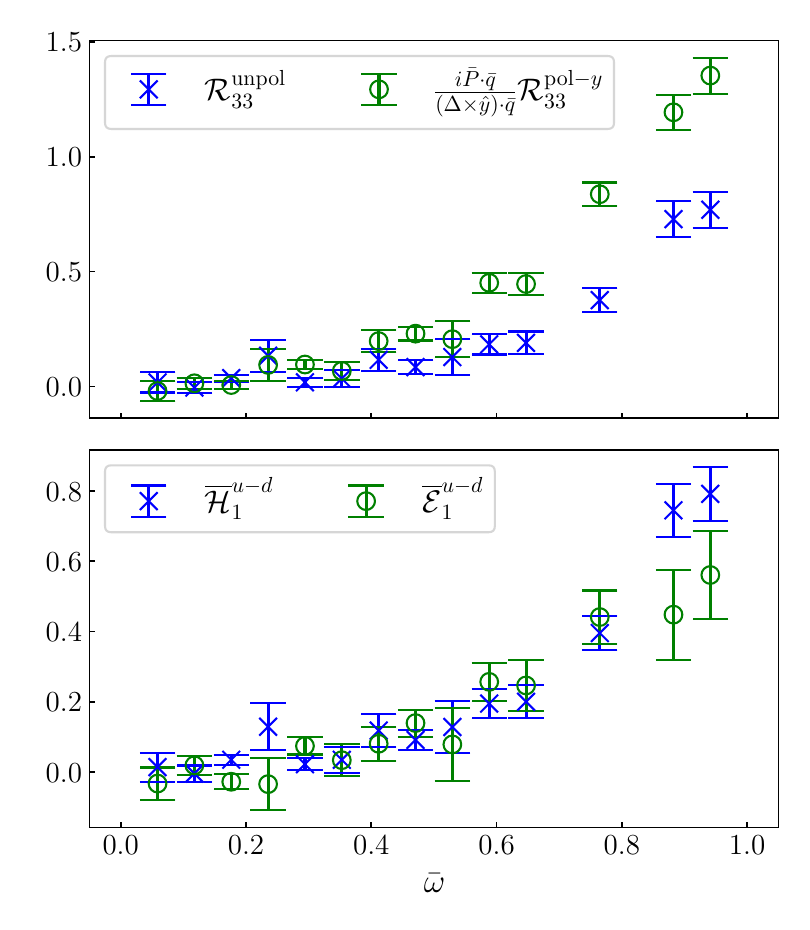}
    \caption{Top: the ratio defined in Eq.~\eqref{spindeplattR} for unpolarised and $y$-direction polarised spin-parity projectors with the $\bomega=0$ term subtracted. Bottom: The resulting subtracted off-forward structure functions, $\overline{\mathcal{H}}_1$ and $\overline{\mathcal{E}}_1$. All results are for $t=-0.57\;\text{GeV}^2$ and $uu-dd$ quarks. }
    \label{hesep}
\end{figure}

\subsection*{Calculation details}

We perform our calculation of the off-forward Compton amplitude on a single set of gauge fields generated by QCDSF/UKQCD \cite{configs} with $2+1$ quark flavours at the SU(3) flavour symmetric point, which yields an unphysical pion mass of $m_{\pi}=412\;\text{MeV}$. See Table \ref{tab:gauge_details} for further details. In calculating the two-point correlator, Eq.~\eqref{pertprop}, we use the nucleon operator,
\begin{equation}
    \chi_\alpha(x) = \varepsilon_{abc} u^a_\alpha(x) \left[ u^b(x) C \gamma_5 d^c(x) \right],
\end{equation}
with the quark fields smeared in a gauge-invariant manner by Jacobi smearing~\cite{Allton:1993wc}, where the smearing parameters are tuned to produce a rms radius of $\simeq 0.5 \, {\rm fm}$.

We determine four sets of perturbed propagators, corresponding to four sets of kinematics for the Compton amplitude with soft momentum transfers $t= 0, -0.29,-0.57,-1.14\;\text{GeV}^2$ all with a hard momentum transfer of $\bar{Q}^2\approx5\;\text{GeV}^2$ (see Table \ref{tab:q_kinematics} for details). Furthermore, we determine the propagators for two magnitudes of the Feynman-Hellmann coupling $\lambda$: for $t=0,-0.29\;\text{GeV}^2$ we use $\lambda=(0.0125, 0.025)$, and for $t=-0.57,-1.14\;\text{GeV}^2$ we use $\lambda=(0.00625, 0.0125)$.

Our determination of the Compton amplitude from these correlators is similar to that from Ref.~\cite{hannafordgunn2021generalised}: we fit our correlator ratio in Eq.~\eqref{FHneat} as a linear function in Euclidean time, using a similar weighted averaging method as in Ref.~\cite{beane_wavg}. Subsequently, we fit this as a quadratic in $\lambda$---see Appendix \ref{sec:appendixdeterm} for details.

For each set of $\mathbf{q}_{1,2}$ we require a new set of inversions---see Eq.~\eqref{q1q2_t_Q2}. However, given a pair of $\mathbf{q}_{1,2}$, the $\bomega$ variable can be expressed as
\begin{align}
    \bomega = \frac{4\mathbf{p}'\cdot(\mathbf{q}_1+\mathbf{q}_2)}{(\mathbf{q}_1+\mathbf{q}_2)^2}
\end{align}
for $\mathbf{\Delta}\cdot\bar{\mathbf{q}}=0$. As such, by varying the sink momentum, $\mathbf{p}'$, we can obtain results at multiple values of $\bomega$ for a single set of $t$ and $\bar{Q}^2$ values. See Appendix \ref{sec:appendixsink}, Tab.~\ref{tab:omega_kinematics_ofca2} for all $\bomega$ values used in this work.

In Fig.~\ref{hesep}, we plot the $\bomega$ dependence of the $t=-0.57\;\text{GeV}^2$ results, and illustrate the determination of the $\mathcal{H}_1$ and $\mathcal{E}_1$ structure functions from the spin-parity traced quantity $\mathcal{R}$, given in Eq.~\eqref{spindeplattR}. We observe good signal for $\mathcal{R}$ for both spin-parity projectors, as well as the subtracted Compton structure functions $\overline{\mathcal{H}}_1$ and $\overline{\mathcal{E}}_1$. Moreover, these quantities all have the $\bomega^2$ polynomial behaviour expected from Eq.~\eqref{subDR}. Results for the $uu$ quark $\overline{\mathcal{H}}_1$ structure function across all $t$ values are presented in Fig.~\ref{w_allfits}; see Fig.~\ref{w_allfits_d} in Appendix \ref{sec:appendixdquark} for $dd$ quark results.

\section{Mellin Moment Fit} \label{sec:moments}

Our first fit strategy is to determine the Mellin moments of the real-time scattering amplitudes. To this end we Taylor expand the dispersion relation Eq.~\eqref{subDR} about $\bomega=0$:
    \begin{align}
   \begin{split}
       \overline{\mathcal{H}}_1(\bomega, t, \bar{Q}^2) & = 2\sum_{n=1}^{\infty}\bomega^{2n}
   \mathcal{A}_{2n,0}(t,\bar{Q}^2),
   \\  \overline{\mathcal{E}}_1(\bomega, t, \bar{Q}^2) & = 2\sum_{n=1}^{\infty}\bomega^{2n}
   \mathcal{B}_{2n,0}(t,\bar{Q}^2),
  \label{momexpansion}
   \end{split}
\end{align}
where 
$\mathcal{A}_{2n,0}$ and $\mathcal{B}_{2n,0}$ are the $n^{\text{th}}$ Mellin moments of the zero-skewness structure functions:
\begin{align}
\begin{split}
    \mathcal{A}_{2n,0}(t,\bar{Q}^2) & = \frac{2}{\pi}\int _0 ^1 dx x^{n-1}\text{Im}\mathcal{H}_1(x, t, \bar{Q}^2),
  \\  \mathcal{B}_{2n,0}(t,\bar{Q}^2) & = \frac{2}{\pi}\int _0 ^1 dx x^{n-1}\text{Im}\mathcal{E}_1(x, t, \bar{Q}^2).
    \label{Mellinmom}
\end{split}
\end{align}
In a twist expansion, these moments can be expressed as \cite{hannafordgunn2021generalised}
\begin{align}
   \begin{split}
        \mathcal{A}_{2n,0}(t,\bar{Q}^2) & = \mathcal{C}\left( \alpha_S(\bar{Q}) \right ) A_{2n,0} (t) + \text{higher twist},
     \\ \mathcal{B}_{2n,0}(t,\bar{Q}^2) & = \mathcal{C}\left( \alpha_S(\bar{Q}) \right ) B_{2n,0} (t) + \text{higher twist},
    \label{twistexpansion}
   \end{split}
\end{align}
where $\mathcal{C}$ is the Wilson coefficient and $A_{n,0}$ and $B_{n,0}$ are the generalised form factors (GFFs), given by the moments of the GPDs $H(x,t)$ and $E(x,t)$, respectively.

We emphasise that Eq.~\eqref{twistexpansion} is included to illustrate the connection between the structure function moments ($\mathcal{A}_{n,0}$ and $\mathcal{B}_{n,0}$) and the leading-twist GFFs (${A}_{n,0}$ and ${B}_{n,0}$). As Eq.~\eqref{twistexpansion} states, it is necessary to project our results to OPE to extract the leading-twist GFFs. In this work, however, we calculate the physical Compton amplitude and extract the moments of the physical structure functions (LHS of Eq.~\eqref{twistexpansion}) akin to an experimental determination, which include the appropriately renormalised leading-twist contributions and the Wilson coefficients that involve the mixing effects, and the higher-twist contributions.

\subsection*{Fit details}

From Eq.~\eqref{momexpansion}, we can fit the leading $N_{\text{max}}$ moments of the subtracted structure functions to a power series in $\bomega$:
\begin{align}
    \begin{split}
        & f(\bomega) =
        2\sum_{n=1}^{N_{\text{max}}}\bomega ^{2n} M_{2n},
        \label{modelindependent_fitfun}
    \end{split}
\end{align}
where $M_{2n}$ are the moments of either $\overline{\mathcal{H}}_1$ or $\overline{\mathcal{E}}_1$ as per Eq.~\eqref{Mellinmom}.

For the $t=-0.29\;\text{GeV}^2$ results we note we can not access the $\bomega=0$ point; see Appendix \ref{sec:appendixsink} for an explanation. As such, we fit the $\bomega=0$ value simultaneously with our moments to the \emph{unsubtracted} structure functions:
\begin{align}
    \begin{split}
        & f(\bomega) =
        f(0)+2\sum_{n=1}^{N_{\text{max}}}\bomega ^{2n} M_{2n}.
        \label{modelindependent_fitfun_t0p29}
    \end{split}
\end{align}

\begin{figure*}[t!] 
\includegraphics[width=0.92\linewidth]{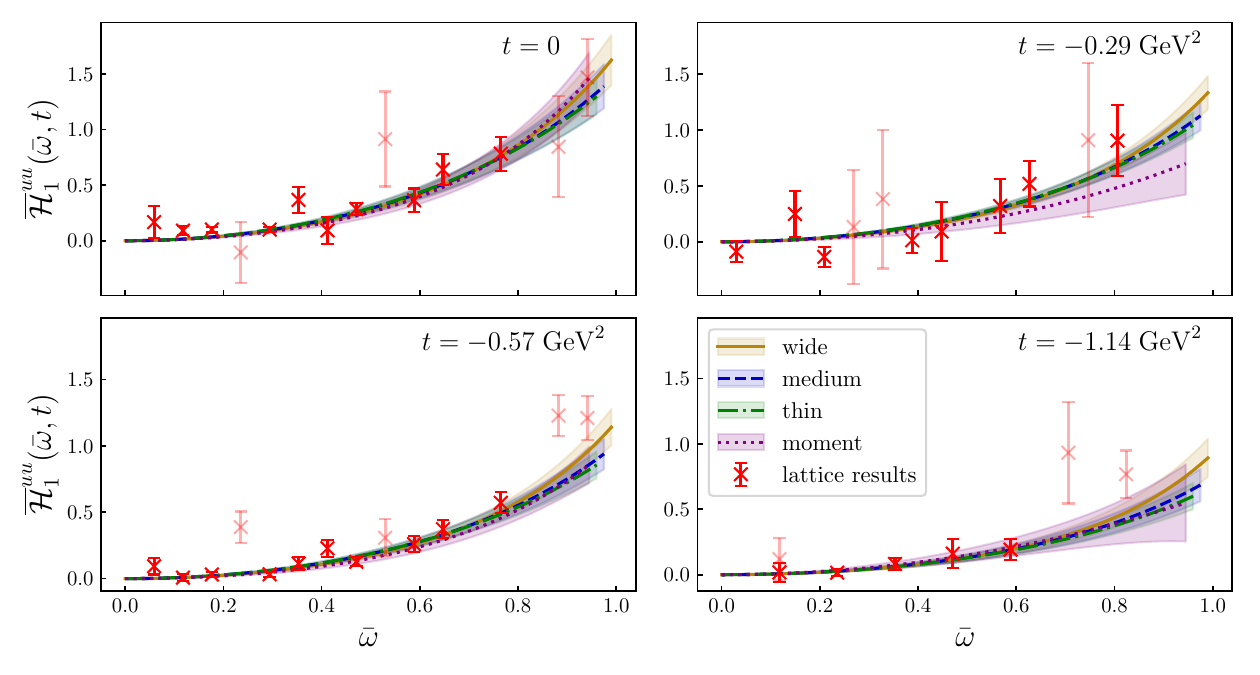}
\caption{The subtracted Compton structure function $\overline{\mathcal{H}}_1(\bomega, t)$ for all $t$ values; $uu$ results only. Note that our fits only use points for which $|\mathbf{p}'|< 1\;\text{GeV}$; the shaded points are those with a sink momentum greater than this. Curves correspond to all the fits performed in this work: the model-independent fit, Eq.~\eqref{modelindependent_fitfun} (`moment'); and the model-dependent fits, Eq.~\eqref{compampmodelparam}, using the three sets of priors in Tab.~\ref{priors_xfit} (`wide', `medium' and `thin'). Note: the upper limit on the $y$-axis is held fixed between the panels to demonstrate the change in magnitude with $-t$.}
\label{w_allfits}
\end{figure*}

This allows us to determine the leading moments of each of the Compton structure functions $\mathcal{H}_1$ and $\mathcal{E}_1$, which at leading-twist are the GFFs $A_{2n,0}$ and $B_{2n,0}$, respectively.

We use the Bayesian Markov chain Monte Carlo (MCMC) package \texttt{PyMC} \cite{pymc3_no1, pymc3_no2} to perform this fit. This allows us to sample the model parameters from prior distributions that reflect physical constraints.

For the $t=0$ kinematics our results are simply the forward Compton structure function, $\mathcal{F}_1$, which is positive definite and hence has monotonically decreasing Mellin moments \cite{fwdpaper}:
\begin{align}
   0 \leq \mathcal{A}_{n+1,0}(t=0) \leq \mathcal{A}_{n,0}(t=0),
\end{align}
where in this section we suppress the $\bar{Q}^2$ argument of the moments for convenience.

Hence for the $n^{\text{th}}$ moment, we use a uniform prior distribution in the range $\mathcal{A}_{n,0}(0)\in[0,\mathcal{A}_{n-2,0}(0)]$. We choose $\mathcal{A}_{2,0}(0)\in[0,1]$ for the prior on the leading moment.

For the off-forward results, we no longer have monotonicity, so we use positivity constraints on the GPDs \cite{Pobylitsa_2002}, which at $\xi=0$ are:
\begin{align}
|H(x,t)|\leq q(x), \quad |E(x,t)|\leq \frac{2m_N}{\sqrt{-t}}q(x),
    \label{GPDpositivity}
\end{align}
where $q(x)$ is the leading-twist parton distribution function. From these, it is simple to determine the bounds on the moments at $\xi=0$:
\begin{align}
 \big|A_{2n,0}(t)\big|\leq  a_{2n}, \quad  \big|B_{2n,0}(t)\big|\leq  \frac{2m_N}{\sqrt{-t}}a_{2n}, 
    \label{priors}
\end{align}
where $a_n$ is the $n^{\text{th}}$ parton distribution function moment.

Although Eq.~\eqref{priors} is derived for leading-twist GPDs we use it nonetheless, noting that we are at a reasonably large $\bar{Q}^2$ and that these bounds are not overly strict. As such, we adapt Eq.~\eqref{priors} to the prior:
\begin{align}
    \big|\mathcal{A}_{2n,0}(t)\big|\leq  \mathcal{A}_{2n,0}(0), \quad  \big|\mathcal{B}_{2n,0}(t)\big|\leq  \frac{2m_N}{\sqrt{-t}}\mathcal{A}_{2n,0}(0).
    \label{offfwdpriors}
\end{align}
For the $\mathcal{A}_{n,0}(0)$ bounds we use the mean plus one standard deviation of the moments calculated from the $t=0$ results.

This fit is performed individually for each $t$ value across the $\bomega$ values given in Tab.~\ref{tab:omega_kinematics_ofca2} of Appendix \ref{sec:appendixsink}. Note that in our fit we only use $\bomega$ values for which the sink momentum is
\begin{align}
    |\mathbf{p}'| < 1\;\text{GeV},
\end{align}
as these are the points for which (1) we can better insure ground state isolation, and (2) $\mathcal{O}(ap_{\mu})$ discretisation artefacts are expected to be negligible for these points. We discuss these systematics further in the next section.

\subsection*{Results}

As we can see in Fig.~\ref{w_allfits}, the moment fit (labelled `moment') describes the $\bomega$-dependence of $\overline{\mathcal{H}}_1$ data well, with most points being consistent within a standard deviation of the fits. Moreover, the fits are even consistent with some of the $|\mathbf{p}'|\geq 1\;\text{GeV}$ points that were not included in the fits. We note that for points with $|\mathbf{p}'|<1\;\text{GeV}$, the $t=0,-0.29,-0.57\;\text{GeV}^2$ results do not have $\bomega$ points greater than $\bomega\approx 0.8$, while for $t=-1.14\;\text{GeV}^2$, there are no such points beyond $\bomega\approx0.6$. This limits our ability to constrain the higher moments in the present work. Moreover, since the $t=-0.29\;\text{GeV}^2$ results require us to fit the $\bomega=0$ subtraction function, instead of determining it directly, these fits are generally of a poorer quality. Note in Fig.~\ref{w_allfits} for the $t=-0.29\;\text{GeV}^2$ we subtract off the fitted $\bomega=0$ point. 

See Appendix \ref{sec:appendixmoments} for the posterior distributions from the Bayesian Markov chain Monte Carlo fits. Note that while using the fit function in Eq.~\eqref{modelindependent_fitfun}, it is necessary to choose the number of moments to fit, $N_{\text{max}}$. To make this choice, we compare the effect of varying this parameter in Fig.~\ref{moment_posteriors}. For $N_{\text{max}}\geq 3$ the order of truncation has negligible impact on the leading moments, and therefore we take $N_{\text{max}}=7$.

\begin{figure}
    \centering
    \includegraphics[width=\linewidth]{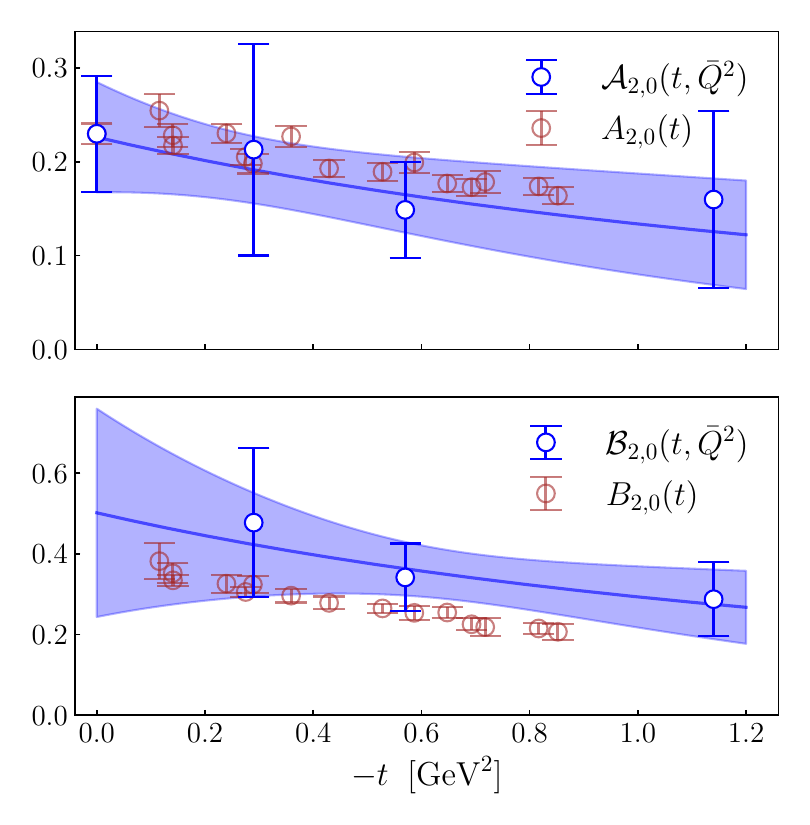}
    \caption{The $n=2$ off-forward Compton amplitude moments $\mathcal{A}_{2n,0}$ (top) and $\mathcal{B}_{2n,0}$ (bottom) determined from a moment fit. These are the moments of $\mathcal{H}_1$ and $\mathcal{E}_1$ as in Eq.~\eqref{Mellinmom}. The Compton amplitude results are for the $uu-dd$ quarks. We compare this to $A_{2,0}$ and $B_{2,0}$ GFFs for $u-d$ quark combination, calculated from the local twist-two lattice operators on the same set of gauge configurations. Note the twist-two operators are renormalised using $\text{RI}^{\prime}\text{/MOM}$., while the OFCA amplitude moments contain all power-corrections and higher-twist contributions.}
    \label{moments}
\end{figure}

In addition to determining our OFCA moments, we also determine the generalised form factors $A_{n,0}$ and $B_{n,0}$ for $n=1,2$ using the local twist-two operators on the same set of gauge configurations. The matrix elements of these local operators are computed using standard three-point function methods\cite{gpdlatt2}---see Appendix \ref{sec:appendix3pt} for the details. As per Eq.~\eqref{twistexpansion}, the off-forward structure function moments $\mathcal{A}_{2n,0}$ and $\mathcal{B}_{2n,0}$ correspond to the GFFs $A_{2n,0}$ and $B_{2n,0}$, respectively, up to higher-twist, power-corrections and the Wilson coefficient. As such, the $A_{2,0}$ and $B_{2,0}$ GFFs determined from local operators are a useful point of comparison.

Finally, we fit $n=2$ moments as a function of $t$, using the dipole parametrisation:
\begin{align}
    G(t) = \frac{G(0)}{\left(1-t/m_{\text{dip}}^2\right)^2}.
    \label{dipole}
\end{align}
Given the large uncertainties on our points, we only use this simple parametrisation and do not test the effects of different parametrisations for form factor fits. See Tab.~\ref{dipole_pars} for the parameters of our dipole fits for $uu-dd$ quarks.

\begin{table}[t!]
    \centering
   \caption{Summary of parameters from dipole fit. All results for $uu-dd$ quarks.}
\label{dipole_pars}
    \begin{tabular}{c|cc}
                   & $G(0)$ & $m_{\text{dip}}$   \\
				\hline
                    \hline
$\mathcal{A}_{2,0}$ &  0.226(59) & 1.8(1.1) \\ 
$\mathcal{B}_{2,0}$ & 0.50(26) & 1.8(1.3) \\ 
    \end{tabular}
\end{table}

In Fig.~\ref{moments} we plot the OFCA moments $\mathcal{A}_{2,0}$ and $\mathcal{B}_{2,0}$ as functions of the soft momentum transfer $t$, including a comparison to ${A}_{2,0}$ and ${B}_{2,0}$ GFFs. We observe good agreement between the helicity-conserving moments $\mathcal{A}_{2,0}$ and the GFF $A_{2,0}$ across the range of $t$ values. Similarly, there is reasonable agreement between the helicity-flipping moments $\mathcal{B}_{2,0}$ and $B_{2,0}$. 

However, we emphasise that, even with complete control of lattice systematics, our structure function moments should be distinct from the leading-twist GFFs, and as such we do not attempt a strong comparison between these results. Nor do we attempt a separation of the leading-twist contributions and the power-corrections; such a determination has been performed on our forward Compton amplitude results where more $Q^2$ values are available \cite{fwdpaperscaling}. An equivalent study of the $\bar{Q}^2$ dependence of the off-forward Compton amplitude could provide useful information on the non-leading-twist contributions to these moments. It is, however, encouraging that there is reasonable agreement between the $t$ dependence of the Compton amplitude moments and that of the local twist-two operators.

Finally, we note that the parameters from our dipole fits broadly agree with other fits to generalised form factors calculated from local twist-two operators at similar pion masses \cite{gpdlatt6}. Moreover, we determine the $u-d$ quark angular momentum from the Ji sum rule \cite{jiog}:
\begin{align}
   \<J_{u-d}^3\> \approx \frac{1}{2}[\mathcal{A}^{uu-dd}_{2,0}(0,\bar{Q}^2)+\mathcal{B}^{uu-dd}_{2,0}(0,\bar{Q}^2)] = 0.36(16),
\end{align}
neglecting the non-leading-twist corrections to our moments. Again, this agrees with determinations from local twist-two operators at similar pion masses \cite{gpdlatt6}, although our errors are very large mostly owing to the statistical uncertainties on the $\mathcal{B}_{2,0}(0,\bar{Q}^2)$ dipole fit. 

Despite the size of these uncertainties, this calculation provides an alternative means of determining the Ji sum rule. Moreover, determinations of the OFCA with multiple $\bar{Q}^2$ values would allow us to analyse the hard scale dependence of this quantity, which is not achievable from other methods.

\section{Model Fit} \label{sec:regge}

In the previous section, we determined the Mellin moments of the real-time off-forward structure functions from our Euclidean OFCA. While this determination is largely model-independent, it is difficult to reconstruct the complete real-time structure functions (and hence GPDs) from a limited set of Mellin moments.

As such, for our second fit strategy we use the phenomenological parametrisation of a GPD (or off-forward structure function), 
\begin{align} 
    H(x,t) = Cx^{-\alpha(t)}(1-x)^{\beta},
    \label{gpdmodelparam}
\end{align}
with $\alpha(t) = \alpha_0 + \alpha' t$, where $\alpha'$ is the Regge slope parameter.

Note that this parametrisation is normalised by the factor
\begin{align}
    C = A\int_0^1 dx x^{-\alpha_0}(1-x)^{\beta} = A\frac{\Gamma(3-\alpha_0+\beta) }{\Gamma(2-\alpha_0)\Gamma(\beta+1)},
\end{align}
which ensures that $\mathcal{A}_{2,0}(t=0,\bar{Q}^2) = A$. This gives us a total of four parameters in our model: $A, \alpha_0, \alpha'$ and $\beta$. We then perform a global fit with this parametrisation to our Compton amplitude results for all $t$ values.

The model in Eq.~\eqref{gpdmodelparam} and similar Regge-inspired parametrisations of GPDs have been used widely to determine GPD properties from various experimental processes \cite{reggeparam1, gpdmodel2004, diehlmodel, FFfromGPD, gpdmodel2006, valenceregge1, reggeparam_dvcs, valenceregge2, globalfitgonzalez, valenceregge3, valenceregge4}, as well as in fits to other lattice results \cite{ji_lattice_phenom_fit, ji_lattice_phenom_fit_xi}.

Inserting the ansatz in Eq.~\eqref{gpdmodelparam} into our dispersion relation, Eq.~\eqref{DR_LT}, we obtain
\begin{align}
    \begin{split}
      &\overline{\mathcal{H}}_1(\bomega, t)= 2C\bomega^2\frac{\Gamma(2-\alpha(t))\Gamma(\beta+1)}{\Gamma(3+\beta-\alpha(t))} 
        \\ & \times\prescript{}{3}F_2\bigg[ {1, (2-\alpha(t))/2,  (3-\alpha(t))/2 \atop  (3+\beta-\alpha(t))/2,  (4+\beta-\alpha(t))/2}; \bomega^2\bigg],
        \label{hypergeom}
    \end{split}
\end{align}
where $\prescript{}{3}F_2$ is a generalised hypergeometric function.

\begin{figure}[t!] 
\includegraphics[width=\linewidth]{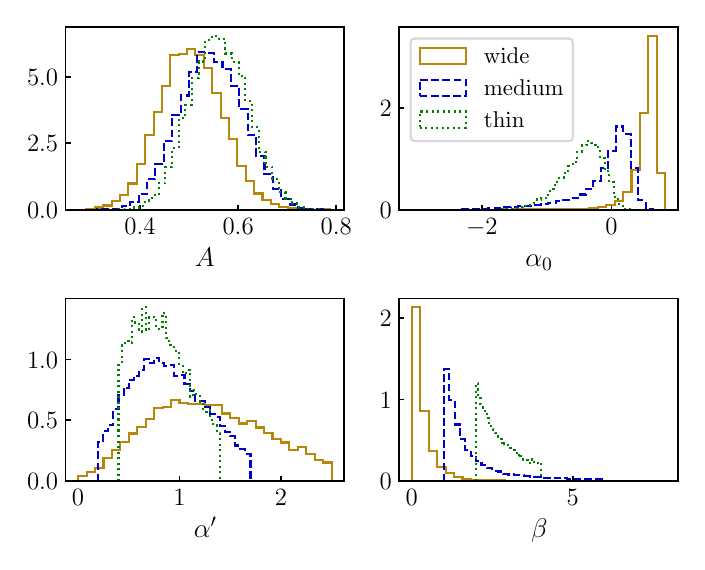}
\caption{Posterior distributions of our three model parameters for $u$ quarks; note that $\alpha_0$ is not a model parameter but its posterior can be reconstructed with Eq.~\eqref{alpha0}.}
\label{posteriors_model}
\end{figure}

Equation \eqref{hypergeom} can be expressed as the sum of moments:
\begin{align}
\overline{\mathcal{H}}_1(\bomega, t) = 2C\sum_{n=1}^{\infty}\bomega^{2n} \frac{\Gamma(2n-\alpha(t))\Gamma(\beta+1)}{\Gamma(1+2n-\alpha(t)+\beta)},
\label{compampmodelparam}
\end{align}
with the $n^{\text{th}}$ moment as
\begin{align}
   \mathcal{A}_{n,0}(t)= C\frac{\Gamma(n-\alpha(t))\Gamma(\beta+1)}{\Gamma(1+n-\alpha(t)+\beta)},
    \label{model_moms}
\end{align}
which is similar to Regge-inspired models of elastic form factors \cite{reggeEFF}. 

To simplify the implementation of the fit, we use Eq.~\eqref{compampmodelparam} as our fit function, truncating at a very high order, $n=50$, which ensures even marginal effects from the higher moments are negligible. 

We note that this model is best justified for valence quarks, though our results include sea quark contributions (i.e.~$H(-x,t)$) that we take to be suppressed. The assumption that our distributions are dominated by valence quark contributions allows us to make the further constraint on our parameters that
\begin{align}
    \int _0 ^1 dx H^q(x,t=0) = F^q_1(|t|=0) = N_q, 
    \label{quarkcounting}
\end{align}
for $N_q$ the number of valence quarks of flavour $q$. While Eq.~\eqref{quarkcounting} is strictly only true for leading-twist, it should nonetheless be a good approximation for $\bar{Q}^2\approx 5\;\text{GeV}^2$.

\begin{figure*}[t!]
            \includegraphics[width=0.92\linewidth]{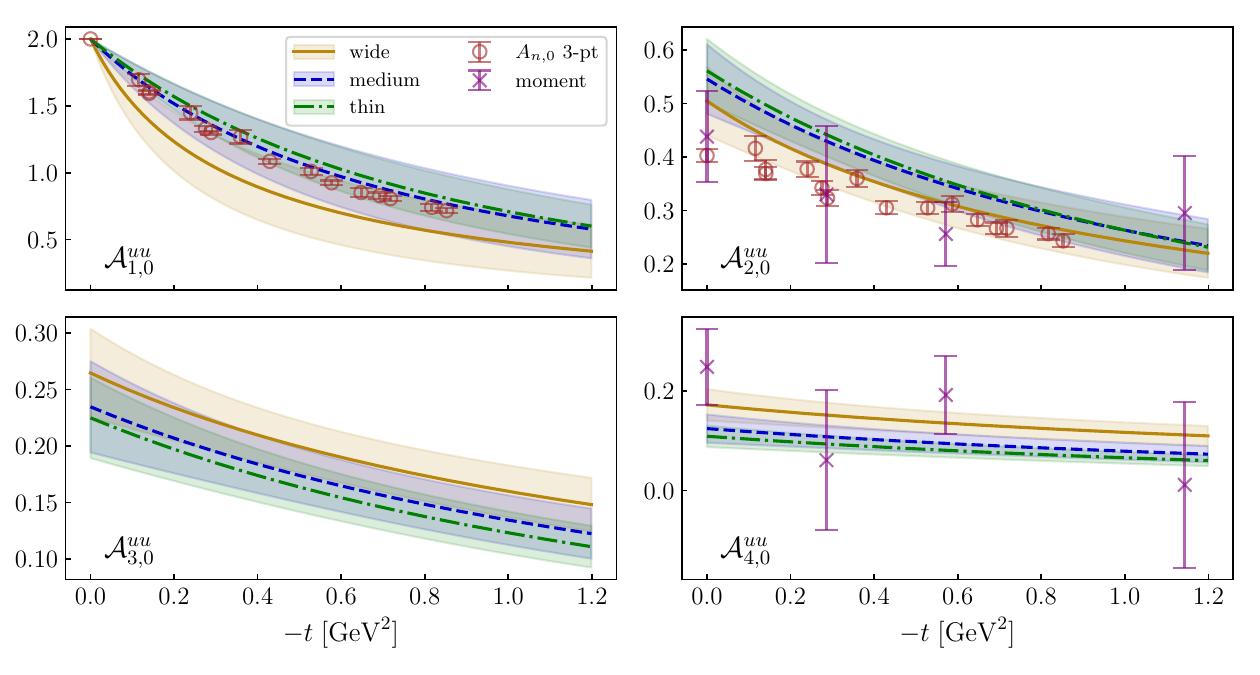}
            \caption{The Compton amplitude moments $\mathcal{A}_{n,0}$, determined from the model-dependent fit, Eq.~\eqref{compampmodelparam} with the three sets of priors (`wide', `medium' and `thin'). We compare these to the moments from the direct moment fit, Eq.~\eqref{modelindependent_fitfun} (`moment'); all results for $u$ quarks. In addition we compare our Compton amplitude moments to the GFFs calculated with local twist-two operators on the same gauge configurations (`$A_{n,0}$ 3-pt'). We emphasise again that the Compton amplitude moments, $\mathcal{A}_{n,0}$ are distinct from the leading-twist GFFs $A_{n,0}$, although they can be related via Eq.~\eqref{twistexpansion}.}
        \label{A_u_allmoms}
        \end{figure*}

Equation \eqref{quarkcounting} gives us
\begin{align}
   \alpha_0 =   \frac{ N_q - A(2 +\beta ) }{N_q - A } ,
   \label{alpha0}
\end{align}
which we use to remove the parameter $\alpha_0$ from our fits.

\subsection*{Fit details}

We fit the model in Eq.~\eqref{compampmodelparam} simultaneously to all our soft momentum transfer values: $t=0,-0.29, -0.57,-1.14\;\text{GeV}^2$. Note that we only fit the $\mathcal{H}_1$ structure function as the $\mathcal{E}_1$ results are typically poorer quality and lack the $t=0$ Compton amplitude.

The fit is again performed with Bayesian MCMC. However, in contrast to the moment fits, there are no model-independent priors for these parameters. As such, to test the dependence of this fit on our priors we vary the width of the priors around approximate phenomenologically expected values: $\alpha'=0.9\;\text{GeV}^{-2}$ \cite{reggeparam1, gpdmodel2004, diehlmodel} and $\beta=3$, while keeping $A\in[0,1]$ for all fits. We fit to our $uu$ and $dd$ structure functions separately, with three parameters for each flavour. All prior distributions are uniform. See Tab.~\ref{priors_xfit} for a summary of the three priors.

As with the Mellin moment fits, we only use $\bomega$ values for which the sink momentum is $|\mathbf{p}'|< 1\;\text{GeV}$.

\begin{table}
    \centering
   \caption{Priors for the three parameters.}
\label{priors_xfit}
    \begin{tabular}{c|ccc}
                   & $A$  & $\alpha'$ [$\text{GeV}^{-2}$]  & $\beta$ \\
                   \hline\hline
                   
     wide    & $[0,1]$   &  $[0.0,2.5]$      &   $[0,8]$  \\
     medium    & $[0,1]$  &  $[0.2,1.7]$      &   $[1,6]$  \\
     thin   & $[0,1]$   &  $[0.4,1.4]$      &   $[2,4]$  
    \end{tabular}
\end{table}

\subsection*{Results}

In Fig.~\ref{w_allfits}, we plot the fits to the $uu$ quark Compton structure function $\overline{\mathcal{H}}_1$ for all $t$ values; see Fig.~\ref{w_allfits_d} for $dd$ quark results. Among the three sets of priors for model-dependent fits, we see good agreement in $\bomega$-space apart from the regions that are not constrained (i.e.~where there are no $\bomega$ values in the fit). Comparing the model-independent moment fit and our model fits, discrepancies are apparent for $\bomega \gtrsim 0.3$.

\begin{table}
\caption{The values of the Regge slope parameter $\alpha'$ for the two flavours and three sets of model priors.}
\label{alpha_res}
    \centering
    \begin{tabular}{c|cc}
                $\alpha'$ $[\text{GeV}^{-2}]$   & $u$ & $d$  \\
                   \hline\hline
    wide & $1.32(57)$ & $1.12(63)$ \\ 
medium & $0.92(36)$ & $0.92(41)$ \\ 
thin    & $0.85(26)$ & $0.89(28)$
    \end{tabular}
\end{table}

In Fig.~\ref{posteriors_model}, we plot the posterior distributions of the model parameters; see Fig.~\ref{posteriors_model_d} of Appendix \ref{sec:appendixdquark} for $d$ quark results. We similarly observe good agreement for the $A$ parameter among the three sets of priors, reasonable agreement in $\alpha'$, and a discrepancy among the three fits for the $\beta$ parameters.

The discrepancies between the model fits for large $\bomega$ and the model-dependence of $\beta$ parameter are related. As discussed in the previous section, higher moments are constrained by the large $\bomega$ values of our Compton amplitude. In terms of our model parametrisation, higher moments are more dependent on the $x\to1$ behaviour of the GPD, which is determine by $\beta$. As such, a unique determination of $\beta$ requires accurate and precise determinations of the Compton amplitude for large $\bomega$, which we currently do not have. We emphasise that the determination of higher-quality large $\bomega$ values is simply a matter of systematic improvement, not a fundamental limitation. 

For the `medium' and `thin' priors, we find $\alpha'$ values of $\approx0.9\;\text{GeV}^{-2}$ for both $u$ and $d$ quarks, while for the `wide' priors the mean of the distributions is slightly larger, although still consistent with the other two fits. This compares well with phenomenological studies, which give the range $\alpha'\approx0.8-1.8\;\text{GeV}^{-2}$ for valence quarks found in both deeply virtual Compton scattering analyses and fits to elastic form factors \cite{reggeparam1, gpdmodel2004, diehlmodel, valenceregge1, globalfitgonzalez, valenceregge3}. Moreover, the `thin' and `medium' results compare well with the value of $\alpha'=0.871(6)\;\text{GeV}^{-2}$ found in global fits to the light hadron masses \cite{regge_traj_global}. See Tab.~\ref{alpha_res} for all $\alpha'$ results.

\begin{figure}[t!]
\includegraphics[width=\linewidth]{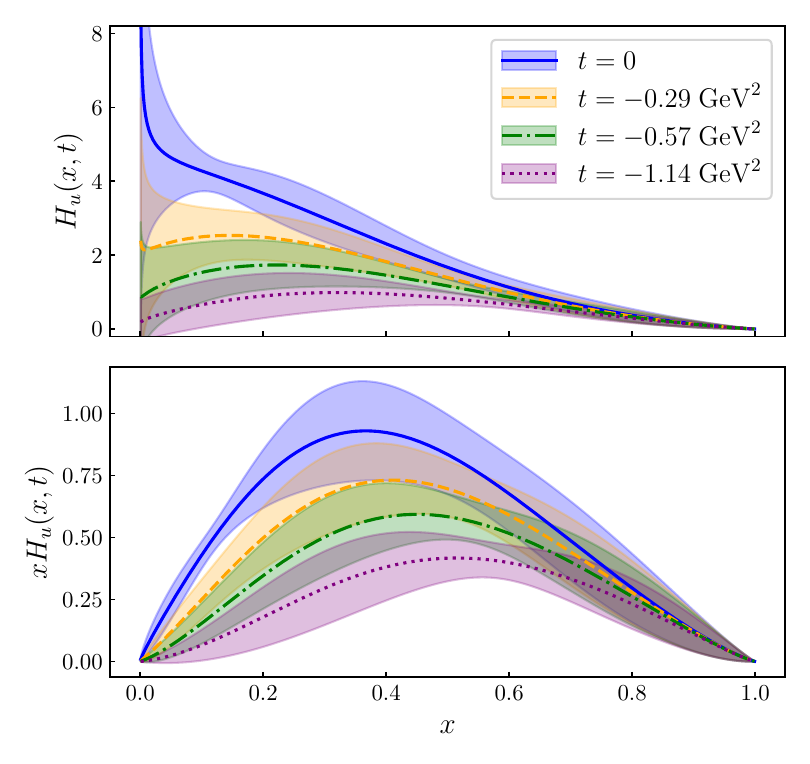}
\caption{The helicity-conserving GPD, $H(x,t)$, for $u$ quarks, determined from fitting the model parametrisation Eq.~\eqref{compampmodelparam} to our Compton amplitude results. This uses the `medium' priors given in Tab.~\ref{priors_xfit}.}
\label{gpd}
\end{figure}

In Fig.~\ref{A_u_allmoms}, we show the $u$ quark results for $\mathcal{A}_{n,0}$, the Mellin moments of $\mathcal{H}_1$ determined from this model fit; see Fig.~\ref{A_d_allmoms} of Appendix \ref{sec:appendixdquark} for $d$ quark results. We also include the moments determined from the model-independent moment fit, Eq.~\eqref{modelindependent_fitfun}. Moreover, we compare these to the Dirac elastic form factor (i.e.~$A_{1,0}$) and $A_{2,0}$ as discussed in the previous section, both determined on the same set of gauge configurations from twist-two local operators---see Appendix \ref{sec:appendix3pt} for details of this calculation. Again, we strongly emphasise, as per Eq.~\eqref{twistexpansion}, our $\mathcal{A}_{n,0}$ moments and the $A_{n,0}$ GFFs are not necessarily equivalent.

We see very strong agreement between our `medium' and `thin' fits and the three-point results for $A_{1,0}$. While this agreement is enforced at $t=0$ by the condition Eq.~\eqref{quarkcounting}, the agreement in the $t$-dependence is nonetheless promising. However, the `wide' parameters show a markedly steeper drop off in $t$. For $\mathcal{A}_{2,0}$ we see a stronger agreement among the three priors for the model fits than for $\mathcal{A}_{1,0}$, and reasonable agreement of these with both the direct moment fits and the three-point results.

\begin{figure}[t!]
\vspace{-4mm}
\includegraphics[width=\linewidth]{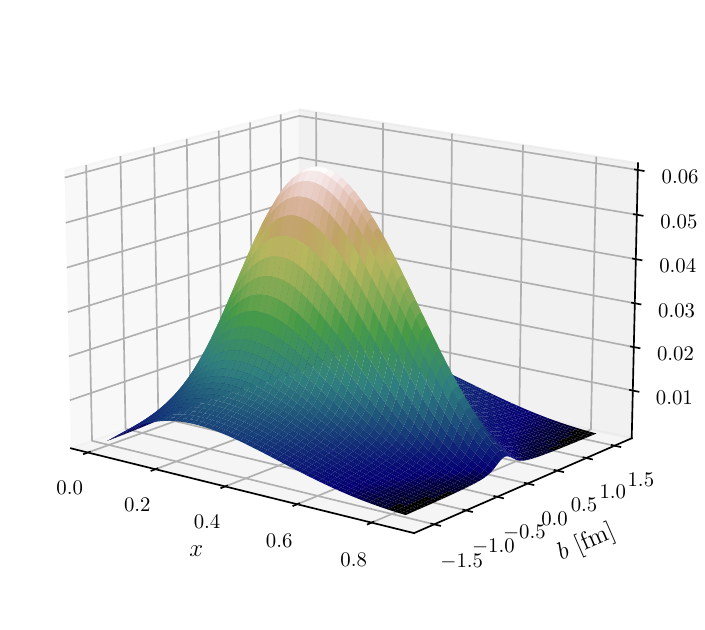}
\caption{The impact parameter space distribution $xH(x,{b})$ for $d$ quarks, determined from the model parametrisation fit using the `medium' priors given in Tab.~\ref{priors_xfit}. Note colours correspond to $z$-axis value, and are simply to help convey the shape.}
\label{ips}
\end{figure}    

For the $\mathcal{A}_{4,0}$ results, there is slightly less agreement among the three sets of priors in the model fit and the direct moment fits are less well-constrained compared to $\mathcal{A}_{2,0}$. As can be seen in Fig.~\ref{w_allfits}, the direct moment fits are highly unconstrained for large $\bomega$, especially for $t=-0.29, -1.14\;\text{GeV}^2$, which explains the irregularity in these results.

In Fig.~\ref{gpd}, we plot the $x$-space GPD for the $u$ quark and `medium' priors. While this result is roughly the expected shape, we note that the GPD appears to have a somewhat slow drop off for $x\to 1$, and the peak of $xH(x,0)$ should be closer to $0.2$ than $0.4$, although the position of this peak is dependent on the hard scale and pion mass. Such features are likely a result of the $\beta$ values skewing small, which is equivalent in moment space to the higher moments not falling quickly enough. As discussed previously, more and better quality results for large $\bomega$ are necessary to better constrain the higher moments and the $\beta$ parameter.

Finally, in Fig.~\ref{ips} we plot the impact parameter space distribution $xH(x,{b})$ \cite{burkardt} obtained via a Fourier transformation ($\mathbf{\Delta}\to\mathbf{b}$, where $b=|\mathbf{b}|$) for the $d$ quark, using the mean values of the `medium' priors. The distribution $H(x,{b})$ can be interpreted as the probability density for a $d$ quark in a fast-moving nucleon as a function of the momentum fraction $x$ and a distance $b$ from the nucleon's centre of mass. Therefore, weighted by $x$, the distribution $xH(x,{b})$ is the momentum density. At this stage given our control of lattice systematics, such a plot is simply a demonstration that our method can be used to reconstruct the impact parameter space distributions.

In general, we note that the GPD ansatz, Eq.~\eqref{gpdmodelparam}, is successful in describing our data and reproducing physically expected properties of GPDs, despite its simplicity. Future studies with higher quality data could test a range of different GPD models.

\subsection*{Systematics and Future Improvements}

The results presented here indicate a great deal of potential for lattice QCD calculations of the off-forward Compton amplitude, as a means to provide unique and complementary physical information about GPDs and off-forward scattering. However, we have also revealed and clarified areas in need of improvement. The most significant of these is the need for more accurate and precise determinations of points with large sink momenta. As discussed, these points give us access to large $\bomega$ values and hence are crucial in constraining higher moments.

To improve the quality of the signal, there exist numerous methods to better isolate the ground state for correlators with large sink momentum \cite{distillation, mom_smearing, adelaide_highmom}, which are already used widely in the calculation of quasi and Ioffe time distributions. Such methods are capable of determining the ground state in correlators with sink momenta as large as $|\mathbf{p}'|\approx 3\;\text{GeV}$, three times greater than the largest sink momentum used in our fits.

In addition, we have treated our $\mathcal{H}_1$ and $\mathcal{E}_1$ data as if they were continuum results, without attempting to account for any discretisation artefacts. We are currently expanding on previous work of a lattice operator product expansion of the Compton amplitude \cite{analytic_wilson_coefficients} that would allow us to control such artefacts.

In Section \ref{sec:regge}, we have adopted a phenomenological parametrisation for our off-forward structure function (Eq.~\eqref{gpdmodelparam}) under the assumptions that the Wilson coefficients are trivial, and the higher-twist effects are negligible at the $\bar Q^2$ scale that we work. While the latter assumption is supported by our forward Compton amplitude calculations~\cite{fwdpaper,fwdpaperscaling}, it would be interesting to resolve the scale dependence of the off-forward structure functions in a future work by studying their $\bar Q^2$ behaviour in a global analysis, similar to what has been attempted for the forward structure functions~\cite{utku_xfit}.

Finally, the lattice systematics that are not specific to our method---the unphysical quark masses, lattice spacing and volume---must be accounted for before a strong comparison can be made with phenomenology.

\section{Summary and Conclusions} \label{sec:summary}

In this study, we present a significantly improved determination of the structure functions of the off-forward Compton amplitude (OFCA). In particular, we determine the structure functions $\mathcal{H}_1$ and $\mathcal{E}_1$ independently for a wider range of kinematics than our previous study \cite{hannafordgunn2021generalised}. This separation allows us to perform a more in-depth attempt at determining properties of the real-time structure functions despite the inverse problem.

We also calculate the $n=1,2$ generalised form factors (GFFs) from the local twist-two lattice operators to compare to our Compton amplitude results. Although we do not perform a perturbative matching of our Compton amplitude moments, we nonetheless note reasonable agreement between the Compton amplitude moments and twist-two GFFs. Similarly, our determinations of the Regge slope parameter $\alpha'$ broadly agree with those from fits to experiment. These agreements are promising, and show that our method is capable of determining meaningful physical information.

Our analysis also clarifies key systematics in need of addressing; in particular, our determinations of the structure functions for large $\bomega$, which requires large sink momenta for our correlators, appear to suffer from lattice artefacts. As discussed in the previous section, addressing these systematics is simply a matter of using and/or building upon existing techniques, making a precise and accurate determination of the off-forward Compton amplitude completely achievable. 

An improved lattice QCD determination of the OFCA would provide a valuable comparison for studies in the quasi- and pseudo-distribution formalisms. Moreover, our method is unique in being able to determine non-leading-twist effects, as has been done for the forward Compton amplitude \cite{comptonproceedings, fwdpaperscaling}. Such effects could provide useful phenomenological information, as most experimental studies of hard exclusive processes have a modest hard scale of $Q^2\approx1-12\;\text{GeV}^2$ and contain additional $|t|/{Q}^2$ corrections \cite{dvcshttheory1, dvcshttheory2}. 

Moreover, past work to determine the subtraction function of the forward Compton amplitude \cite{interlacingsubtraction} could be extended to the OFCA subtraction function. This off-forward subtraction function is a key input for determinations of the proton pressure distribution \cite{dtermexp, pressuredistcomment}, and as such could significantly reduce the errors of model-independent measurements of this quantity.

\section*{Acknowledgements} \label{sec:acknowledgements}

The numerical configuration generation (using the BQCD lattice QCD program~\cite{Haar:2017ubh})) and data analysis (using the Chroma software library~\cite{Edwards:2004sx}) was carried out on the DiRAC Blue Gene Q and Extreme Scaling (EPCC, Edinburgh, UK) and Data Intensive (Cambridge, UK) services, the GCS supercomputers JUQUEEN and JUWELS (NIC, J\"{u}lich, Germany) and resources provided by HLRN (The North-German Supercomputer Alliance), the NCI National Facility in Canberra, Australia (supported by the Australian Commonwealth Government) and the Phoenix HPC service (University of Adelaide). AHG and JAC are supported by an Australian Government Research Training Program (RTP) Scholarship. RH is supported by STFC through grants ST/T000600/1 and ST/X000494/1. PELR is supported in part by the STFC under contract ST/G00062X/1. GS is supported by DFG Grant No. SCHI 179/8-1. KUC, RDY and JMZ are supported by the Australian Research Council grants DP190100297 and DP220103098.

\appendix

\onecolumngrid

\section{Isolating \texorpdfstring{$\mathcal{H}_1$}{H1} and \texorpdfstring{$\mathcal{E}_1$}{E1}} 
\label{sec:heisolation}

We start with the tensor decomposition from Ref.~\cite{hannafordgunn2021generalised}, removing the terms that vanish for $\xi=0$ \cite{tarrach}:
\begin{equation}
    \begin{split}
        & \bar{T}_{\mu\nu}  = \frac{1}{2\pq}\bigg\{-\Big ( h\cdot \bar{q} \mathcal{H}_1+ e\cdot \bar{q} \mathcal{E}_1  \Big)g_{\mu\nu} + \frac{1}{\pq}\Big ( h\cdot \bar{q} \mathcal{H}_2+ e\cdot \bar{q} \mathcal{E}_2  \Big)\bar{P}_{\mu}\bar{P}_{\nu} + \mathcal{H}_3 h_{\{\mu}\bar{P}_{\nu\}}\bigg \}
        \\ &  + \frac{i}{2\pq}\epsilon_{\mu\nu\rho\kappa}\bar{q}^{\rho}\bigg\{ \tilde{h}^{\kappa}  \tilde{\mathcal{H}}_1+ \tilde{e}^{\kappa} \tilde{\mathcal{E}}_1   +\frac{1}{\pq}\Big [ \big (\pq \tilde{h}^{\kappa}-\tilde{h}\cdot \bar{q}\bar{P}^{\kappa}\big) \tilde{\mathcal{H}}_2  \Big] \bigg\} + \Big(\bar{P}_{\mu}q'_{\nu}+\bar{P}_{\nu}q_{\mu}\Big) \Big( h\cdot \bar{q} \mathcal{K}_1+ e\cdot \bar{q} \mathcal{K}_2 \Big) 
        \\ &  + q_{\mu}q'_{\nu}\big(h\cdot\bar{q}-e\cdot\bar{q}\big)\mathcal{K}_5  + \Big(h_{\mu}q'_{\nu}+h_{\nu}q_{\mu}\Big)\mathcal{K}_7 , 
        \label{expltensordecomp2}
    \end{split}
\end{equation}
where we note $\bar{T}_{\mu\nu}$ is the Compton amplitude without gauge projection; that is, it contains no terms with uncontracted $q'_{\mu}$ and $q_{\nu}$. Further note that $h_{\mu}$ and $e_{\mu}$ are given in Eq.~\eqref{hebilineardef} and $\tilde{h}_{\mu} = \bar{u}(P')\gamma_{\mu}\gamma_5u(P)$ and $\tilde{e}_{\mu} = \Delta_{\mu}\bar{u}(P')\gamma_5u(P)/2m_N$.

The key kinematic choice for this work in contrast to Ref.~\cite{hannafordgunn2021generalised}, is that we take $\mathbf{\hat{e}}_k\propto\mathbf{\Delta} $, where $\mathbf{\hat{e}}_k$ is the vector that picks the direction of the current in Eq.~\eqref{pert_mat}. As both $\mathbf{\bar{q}}$ and $\mathbf{\bar{p}}$ are orthogonal to $\mathbf{\Delta}$, the choice $\mathbf{\hat{e}}_k \propto \mathbf{\Delta} $ means that any terms in our tensor decomposition with an uncontracted $\bar{q}_{\mu}$ or $\bar{P}_{\mu}$ vanish. Therefore, only tensor structures with $g_{\mu\nu}$ or $\Delta_{\mu}\Delta_{\nu}$ survive. The former are associated with $\mathcal{H}_1$ and $\mathcal{E}_1$ amplitude, while the latter are suppressed.

As such, this kinematic choice minimises the effects of EM gauge dependent terms (i.e.~any terms with uncontracted $q'_{\mu}$ and $q_{\nu}$) and hence discretisation artefacts, as our local current does not satisfy the continuum Ward identities \cite{latticeWI1, latticeWI2}. Moreover, it helps us isolate $\mathcal{H}_1$ and $\mathcal{E}_1$ instead of a linear combination of other structure functions.

Explicitly, this choice means that the polarised structure functions, $\tilde{\mathcal{H}}_{1,2}$ and $\tilde{\mathcal{E}}_{1,2}$, are attached to
\begin{equation*}
    \Delta_{\{\mu}\epsilon_{\nu\}\sigma\rho\kappa}\Delta^{\sigma}\bar{q}^{\rho}\tilde{h}^{\kappa},
\end{equation*}
after gauge projection. Since $\mu, \nu$ are the Compton amplitude's indices, with the $\mathbf{\hat{e}}_k\propto\mathbf{\Delta} $ condition we have $\mu=\nu=\sigma$, and hence the above equation must vanish. This completely removes all polarised amplitudes.

Further, the $\mathcal{K}_{1,2,5,7}$ amplitudes have no $g_{\mu\nu}$ tensor structure. Therefore, after gauge projection, the only contribution that survives is
\begin{equation*}
    \frac{\Delta_k\Delta_k}{q\cdot q'}\mathcal{K}_{1,2,5,7} \sim \frac{-t}{\bar{Q}^2}\mathcal{K}_{1,2,5,7}.
\end{equation*}
Hence these tensor structures, which are already non-leading-twist, receive an additional kinematic suppression of $|t|/\bar{Q}^2$.

Therefore, up to highly suppressed terms containing the $\mathcal{K}$ amplitudes, the gauge-projected Compton amplitude is
\begin{equation}
    \begin{split}
       & {T}_{kk}  = \frac{1}{2\pq}\bigg[\Big ( h\cdot \bar{q} \mathcal{H}_1+ e\cdot \bar{q} \mathcal{E}_1  \Big)\bigg(1+\frac{q'_kq_k}{q\cdot q'} \bigg) + \frac{1}{\pq}\Big ( h\cdot \bar{q} \mathcal{H}_2+ e\cdot \bar{q} \mathcal{E}_2  \Big)
        \\ &\times \bigg(\bar{P}_k\bar{P}_k-\frac{\pq}{q\cdot q'}(q'_k\bar{P}_k+\bar{P}_kq_k)+ \bigg(\frac{\pq}{q\cdot q'}\bigg)^2q'_kq_k \bigg)  + \mathcal{H}_3 \bigg(\bar{P}_kh_k-\frac{\pq}{q\cdot q'}(q'_kh_k+h_kq_k)+ \frac{\pq h\cdot\bar{q}}{(q\cdot q')^2}q'_kq_k \bigg) \bigg ].
        \label{tensordecompgauge}
        \end{split}
        \end{equation}
       The Dirac bilinear $h_{\mu}$ is orthogonal to $\Delta_{\mu}$, so that the $h_k$ terms, which are proportional to $h\cdot \Delta$, must vanish. Moreover, as previously explained, $\bar{P}_k = 0 = \bar{q}_k$. Therefore, Eq.~\eqref{tensordecompgauge} becomes
        \begin{equation}
    \begin{split}
        {T}_{kk} & = \frac{1}{2\pq}\bigg[\Big ( h\cdot \bar{q} \mathcal{H}_1+ e\cdot \bar{q} \mathcal{E}_1  \Big)\bigg(1-\frac{\Delta_k\Delta_k}{4q\cdot q'} \bigg)
         - \frac{\pq}{4(q\cdot q')^2}\Big ( h\cdot \bar{q} (\mathcal{H}_2+\mathcal{H}_3)+ e\cdot \bar{q} \mathcal{E}_2  \Big){\Delta_k\Delta_k} \bigg ].
        \label{tensordecompgauge2}
        \end{split}
        \end{equation}
        In Ref.~\cite{hannafordgunn2021generalised} it was shown that for large $\bar{Q}^2$ the $\mathcal{H}$ and $\mathcal{E}$ structure functions satisfy the off-forward Callan-Gross relation:
        \begin{equation}
            \frac{\bomega}{2}\big(\mathcal{H}_2 + \mathcal{H}_3 \big) = \mathcal{H}_1 + \Delta \mathcal{H}_{\text{CG}}, \quad \frac{\bomega}{2}\mathcal{E}_2 = \mathcal{E}_1+ \Delta \mathcal{E}_{\text{CG}},
        \end{equation}
        where we have included $\Delta \mathcal{H}_{\text{CG}}$ and $\Delta \mathcal{E}_{\text{CG}}$, the $\mathcal{O}(\alpha_S)$ violations to this Callan-Gross relation.
        
        Further, note that $q\cdot q' = -\bar{Q}^2+t/4$. Hence Eq.~\eqref{tensordecompgauge2} becomes
        \begin{equation}
    \begin{split}
        {T}_{kk} & = \frac{1}{2\pq}\bigg\{ \Big ( h\cdot \bar{q} \mathcal{H}_1+ e\cdot \bar{q} \mathcal{E}_1 \Big)\bigg(1-\frac{\Delta_k\Delta_k}{4(\bar{Q}^2-t/4)}\frac{t}{4(\bar{Q}^2-t/4)} \bigg)
         \\ & - \Big ( h\cdot \bar{q} \Delta\mathcal{H}_{\text{CG}}+ e\cdot \bar{q} \Delta\mathcal{E}_{\text{CG}} \Big)\frac{\Delta_k\Delta_k}{4(\bar{Q}^2-t/4)}\frac{\bar{Q}^2}{\bar{Q}^2-t/4} 
         \bigg \}.
        \label{tensordecompgauge3}
        \end{split}
        \end{equation}
        Given that $\Delta\mathcal{H}_{\text{CG}}$ and $\Delta\mathcal{E}_{\text{CG}}$ are $\mathcal{O}(\alpha_S)$, with the extra $|t|/\bar{Q}^2$ suppression, they are at best $\bar{Q}^{-3}$. Therefore, up to $\bar{Q}^{-3}$ corrections, the OFCA is
        \begin{equation}
    \begin{split}
        {T}_{kk} & = \frac{1}{2\pq}\Big ( h\cdot \bar{q} \mathcal{H}_1+ e\cdot \bar{q} \mathcal{E}_1 \Big).
        \label{finalofca2tensor}
        \end{split}
        \end{equation}
       This is a drastic improvement on Ref.~\cite{hannafordgunn2021generalised}, where we truncated all terms that were not leading-order ($\bar{Q}^{-1}$ and higher), and only isolated a linear combination of $\mathcal{H}_{1,2,3}$ and $\mathcal{E}_{1,2}$. Here, we have either eliminated completely unwanted tensor structures, or suppressed them by a further $|t|/\bar{Q}^2$, with only a simple linear combination of $\mathcal{H}_{1}$ and $\mathcal{E}_{1}$ remaining.

\section{Determination of the Compton amplitude} \label{sec:appendixdeterm}

As in Eq.~\eqref{FHneat}, we determine the combination of perturbed correlators defined in Eq.~\eqref{combocorr}:
\begin{align*}
    R_{\lambda} = \frac{\mathcal{G}_{(\lambda,\lambda)}+\mathcal{G}_{(-\lambda,-\lambda)}-\mathcal{G}_{(\lambda,-\lambda)}-\mathcal{G}_{(-\lambda,\lambda)}}{\mathcal{G}_{(0,0)}}.
\end{align*}
This combination isolates the $\lambda^2$ contribution up to $\mathcal{O}(\lambda^4)$ corrections. 

First, we fit $R_{\lambda}$ as a function of the Euclidean time $\tau$. Apropos Eq.~\eqref{FHneat}, these correlators should have a linear $\tau$ dependence, so we fit the function $f(\tau) = A\tau + B$, where the slope is proportional to the Compton amplitude. This fit is performed using a weighted averaging method similar to Ref.~\cite{beane_wavg}, where multiple Euclidean time windows are fit and then averaged over with each window weighted by
    \begin{equation}
    \tilde{w}^i = \frac{p(\delta A^i)^{-2}}{\sum_{i'} p(\delta A^{i'})^{-2}},
    \label{beaneweight}
\end{equation}
    where $A^i$ is the slope parameter from the $i^{\text{th}}$ fit window, $\delta A^i$ is the statistical error and $p$ is the $p$-value determined by
    $$p = \tilde{\Gamma}(N_{\text{dof}}/2, \chi^2/2)/\tilde{\Gamma}(N_{\text{dof}}/2),$$ 
    where $\tilde{\Gamma}$ the regularised upper incomplete gamma function.

\begin{figure*}[t!] 
\includegraphics[width=0.92\linewidth]{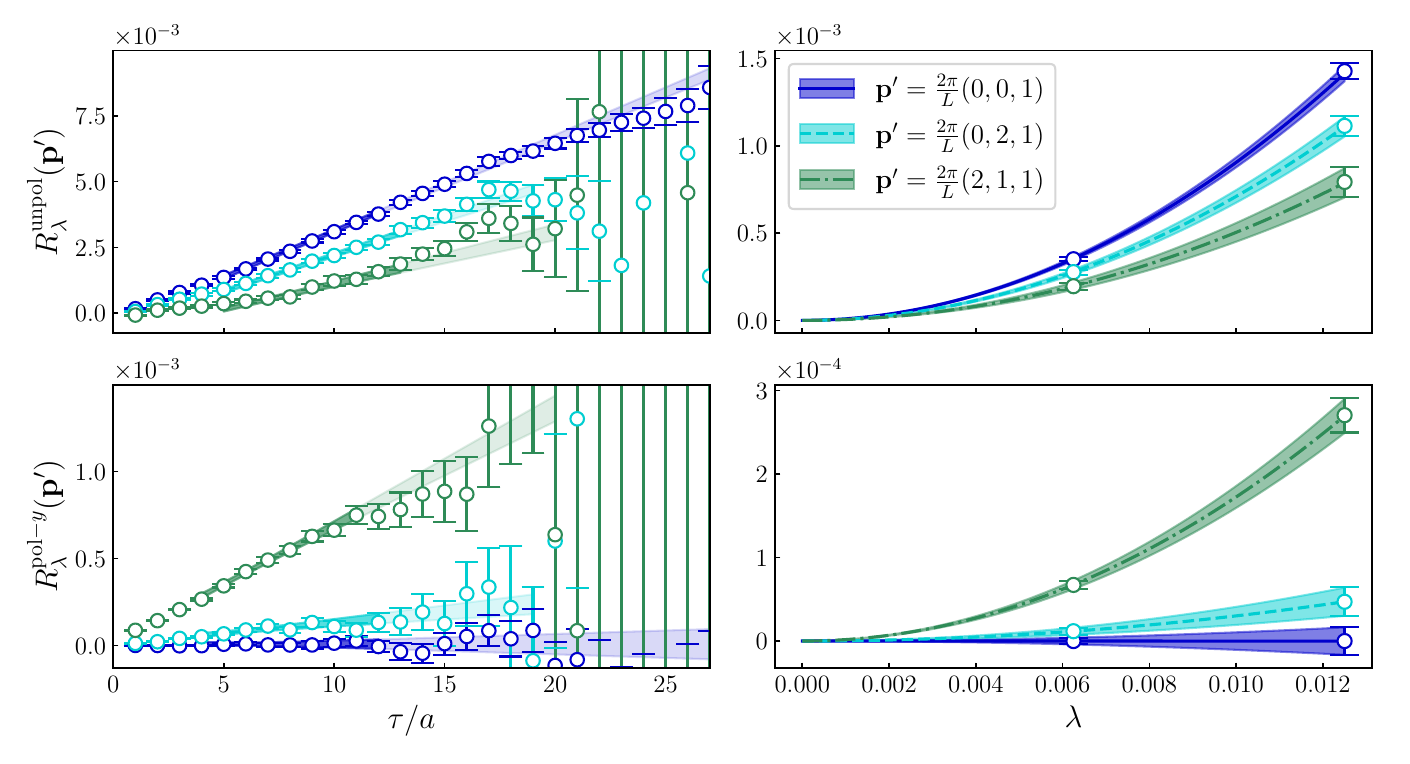}
\caption{Left: the fits in $\tau$ to the quantity $R_{\lambda}$ defined in Eq.~\eqref{combocorr}. The light shaded band is the full extent of all fit windows, while the darker shaded band is the fit window with the highest weight given by Eq.~\eqref{beaneweight}. Right: the fits in $\lambda$ to $R_{\lambda}$ after the fits in Euclidean time. The top panels are for the unpolarised spin-parity projector, while the bottom are for the $y$-polarised projector. All results for $t=-0.57\;\text{GeV}^2$ and $uu$ quark combination.}
\label{lam_tau_fits}
\end{figure*}    

    The minimum Euclidean time to include in the fits is chosen by eye, while the largest is chosen using the unperturbed correlator's noise-to-signal (we choose a number of standard deviations, $N_{\sigma}$, and the largest time slice is chosen as last time slice for which the unperturbed correlator is more than $N_{\sigma}$ standard deviations from zero). Finally, to keep the degrees of freedom greater than zero, we need a fit window of minimum length $3a$.

    After the Euclidean fits, we perform fits in the Feynman-Hellmann parameter $\lambda$. The combination of correlators $R_{\lambda}$ must be a polynomial in $\lambda^2$, where the $\lambda^2$ coefficient is proportional to the Compton amplitude. Since we only determine two $\lambda$ values in the range $[0.0625,0.025]$ for each of our inversions, we only perform a one parameter fit: $f(\lambda) = b\lambda^2$. However, as tested in Ref.~\cite{hannafordgunn2021generalised} the higher order in $\lambda$ terms do not have a major impact on our results, especially at our current level of precision.

    The results of these fits for $t=-0.57\;\text{GeV}^2$ and $uu$ quarks are presented in Fig.~\ref{lam_tau_fits}. We observe that in Euclidean time the results, while still reasonably noisy, are well described by the linear fit function. Similarly, in the Feynman-Hellmann parameter, the results are well-described by the quadratic fit.

\section{Sink momenta}
\label{sec:appendixsink}

In our Compton amplitude determination, the sink momentum determines the value of $\bomega$, which encodes the $x$-dependence of the GPD. In particular, accurate and precise determinations of large $\bomega$ values are crucial to determining higher moments and allowing a full reconstruction of the GPD.

It is convenient to define the dimensionless sink momentum:
\begin{equation}
    \mathbf{n}' = \left ( \frac{L}{2\pi} \right ) \mathbf{p}',
    \label{dimlesssink}
\end{equation}
where for our work the momentum interval is $2\pi/L\approx 380\;\text{MeV}$. 

For the forward results we do not calculate the Compton amplitude for $\mathbf{n}'^2>9$, while for the off-forward Compton amplitude we do not calculate beyond $\mathbf{n}'^2>10$. Without methods to improve the isolation of the ground state, beyond these sink momenta the signal is of poor quality. As per our discussion in Section \ref{sec:bckgrnd}, we do not calculate results for $|\bomega|>1$.

Similarly, in all our fits we do not include results for which $\mathbf{n}'^2\geq 7$, which corresponds to $|\mathbf{p}'|\geq 1\;\text{GeV}$, as ground state isolation is generally poorer and $\mathcal{O}(ap_{\mu})$ artefacts are expected to be significant for these sink momenta. This is discussed further in the text.

\begin{table*}[t!]
\caption{\label{tab:t2t4} Dimensionless sink momenta, $\mathbf{n}'$, and corresponding $\bomega$ values for the four different sets of Compton amplitude results. Momenta in italics have $|\mathbf{p}'|\geq 1\;\text{GeV}$ and are hence excluded from our fits.}
\vspace{2mm}
    \centering
	\begin{tabular}{ccc}
 \begin{tabular}[t]{ccc}
 \multicolumn{3}{c}{$t=0$} \\
		\hline\hline
			 $\mathbf{n}' $ & $\bomega$ & $\mathbf{n}'^2$\\
			\hline\hline
 $(0,0,0)$ & $0.0$ & $0$ \\  
$(-1,2,0)$ & $0.06$ & $5$ \\  
$(1,-1,0)$ & $0.12$ & $2$ \\  
$(0,1,0)$ & $0.18$ & $1$ \\  
$\mathit{(2,-2,0)}$ & $\mathit{0.24}$ & $\mathit{8}$ \\  
$(1,0,0)$ & $0.29$ & $1$ \\  
$(0,2,0)$ & $0.35$ & $4$ \\  
$(2,-1,0)$ & $0.41$ & $5$ \\  
$(1,1,0)$ & $0.47$ & $2$ \\  
$\mathit{(0,3,0)}$ & $\mathit{0.53}$ & $\mathit{9}$ \\  
$(2,0,0)$ & $0.59$ & $4$ \\  
$(1,2,0)$ & $0.65$ & $5$ \\  
$(2,1,0)$ & $0.76$ & $5$ \\  
$\mathit{(3,0,0)}$ & $\mathit{0.88}$ & $\mathit{9}$ \\  
$\mathit{(2,2,0)}$ & $\mathit{0.94}$ & $\mathit{8}$ \\
			\hline\hline
		\end{tabular}
		\quad 
		\begin{tabular}[t]{ccc}
   \multicolumn{3}{c}{$t=-0.29\;\text{GeV}^2$} \\
		\hline\hline
			 $\mathbf{n}'$ & $\bomega$ & $\mathbf{n}'^2$\\
			\hline\hline
$(1,0,-1)$ & $0.03$ & $2$ \\  
$(0,-1,2)$ & $0.15$ & $5$ \\  
$(1,0,0)$ & $0.21$ & $1$ \\  
$\mathit{(2,1,-2)}$ & $\mathit{0.27}$ & $\mathit{9}$ \\  
$\mathit{(0,-1,3)}$ & $\mathit{0.33}$ & $\mathit{10}$ \\
$(1,0,1)$ & $0.39$ & $2$ \\  
$(2,1,-1)$ & $0.45$ & $6$ \\  
$(1,0,2)$ & $0.57$ & $5$ \\ 
$(2,1,0)$ & $0.63$ & $5$ \\  
$\mathit{(1,0,3)}$ & $\mathit{0.75}$ & $\mathit{10}$ \\
$(2,1,1)$ & $0.81$ & $6$ \\
			\hline\hline
		\end{tabular}
		\quad 
		\begin{tabular}[t]{ccc}
   \multicolumn{3}{c}{$t=-0.57\;\text{GeV}^2$} \\
		\hline\hline
			 $\mathbf{n}' $ & $\bomega$ & $\mathbf{n}'^2$\\
			\hline\hline
 $(0,0,1)$ & $0.0$ & $1$ \\  
$(-1,2,1)$ & $0.06$ & $6$ \\  
$(1,-1,1)$ & $0.12$ & $3$ \\  
$(0,1,1)$ & $0.18$ & $2$ \\  
$\mathit{(2,-2,1)}$ & $\mathit{0.24}$ & $\mathit{9}$ \\  
$(1,0,1)$ & $0.29$ & $2$ \\  
$(0,2,1)$ & $0.35$ & $5$ \\  
$(2,-1,1)$ & $0.41$ & $6$ \\  
$(1,1,1)$ & $0.47$ & $3$ \\  
$\mathit{(0,3,1)}$ & $\mathit{0.53}$ & $\mathit{10}$ \\  
$(2,0,1)$ & $0.59$ & $5$ \\  
$(1,2,1)$ & $0.65$ & $6$ \\  
$(2,1,1)$ & $0.76$ & $6$ \\  
$\mathit{(3,0,1)}$ & $\mathit{0.88}$ & $\mathit{10}$ \\  
$\mathit{(2,2,1)}$ & $\mathit{0.94}$ & $\mathit{9}$ \\
			\hline\hline
		\end{tabular}
		\quad
			\begin{tabular}[t]{ccc}
    \multicolumn{3}{c}{$t=-1.14\;\text{GeV}^2$} \\
		\hline\hline
			 $\mathbf{n}'$ & $\bomega$ & $\mathbf{n}'^2$\\
		\hline\hline
$(1,-1,0)$ & $0.0$ & $2$ \\  
$(2,0,-1)$ & $0.12$ & $5$ \\  
$\mathit{(0,-2,2)}$ & $\mathit{0.12}$ & $\mathit{8}$ \\  
$(1,-1,1)$ & $0.24$ & $3$ \\  
$(2,0,0)$ & $0.35$ & $4$ \\  
$(1,-1,2)$ & $0.47$ & $6$ \\  
$(2,0,1)$ & $0.59$ & $5$ \\  
$\mathit{(3,1,0)}$ & $\mathit{0.71}$ & $\mathit{10}$ \\  
$\mathit{(2,0,2)}$ & $\mathit{0.82}$ & $\mathit{8}$ \\ 
			\hline\hline
		\end{tabular}
	\end{tabular}
    \label{tab:omega_kinematics_ofca2}
\end{table*}

In Tab.~\ref{tab:omega_kinematics_ofca2} we present the dimensionless sink momentum and corresponding $\bomega$ values for all results in this work. Note that we average over equivalent kinematics: for the unpolarised projector, the value of $\mathcal{R}_{\mu\mu}$ (see Eq.~\eqref{spindeplattR}) does not change with $\bomega\to-\bomega$ or $\Delta\to-\Delta$, and hence we average over these. For the polarised projector, there is a relative minus sign for each of the changes $\bomega\to-\bomega$ or $\Delta\to-\Delta$ in $\mathcal{R}_{\mu\mu}$, and hence we average over these accounting for the minus sign.

As mentioned in the text, there is no $\bomega=0$ point for the $t=-0.29\;\text{GeV}^2$ results. Recall from Eq.~\eqref{q1q2_t_Q2} that $\bomega\propto \bar{\mathbf{p}} \cdot \bar{\mathbf{q}}$, where 
\begin{align}
    \bar{\mathbf{p}} = \frac{1}{2}({\mathbf{p}}' +{\mathbf{p}}) = \frac{1}{2}(2{\mathbf{p}}' +{\mathbf{\Delta}}).
\end{align}
Therefore, to access $\bomega=0$ we need to set $\bar{\mathbf{p}} = \mathbf{0}$, which implies $\mathbf{p}' =-{\mathbf{\Delta}}/2$ from the above equation. However, for the $t=-0.29\;\text{GeV}^2$ results, $\mathbf{\Delta}=\frac{2\pi}{L}(1,-1,0)$, and hence $\bomega=0$ implies that $\mathbf{p}' =\frac{2\pi}{L}(-0.5,0.5,0)$, which is not accessible with our discretised momentum.

\section{Additional \texorpdfstring{$d$}{d} quark results} \label{sec:appendixdquark}

\begin{figure*}[t!] 
\includegraphics[width=0.92\linewidth]{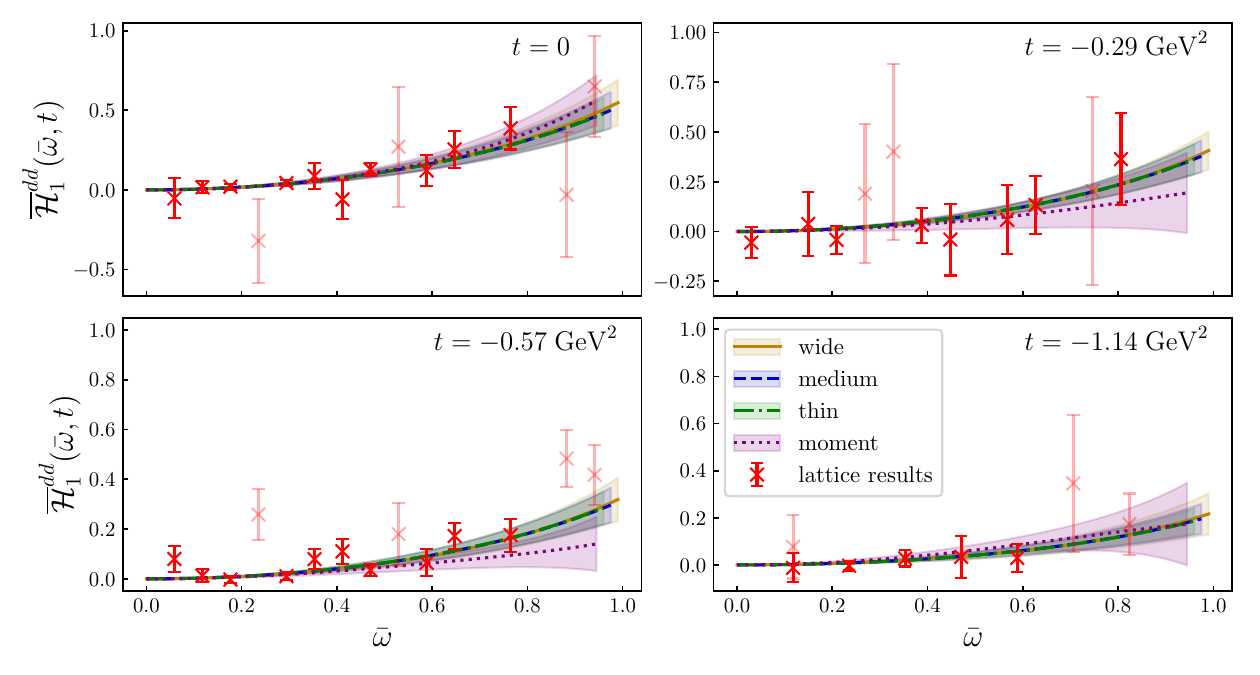}
\caption{All results and labels as in Fig.~\ref{w_allfits} except for $d$ quarks.}
\label{w_allfits_d}
\end{figure*}

In Fig.~\ref{w_allfits_d} we present all the fits performed in $\bomega$ space for the $d$ quarks for the $\mathcal{H}_1$ structure function. We note that the $d$ quark results typically have a poorer signal-to-noise ratio than those for the $uu$ quark combination.

\begin{figure*}[t!] 
\includegraphics[width=0.9\linewidth]{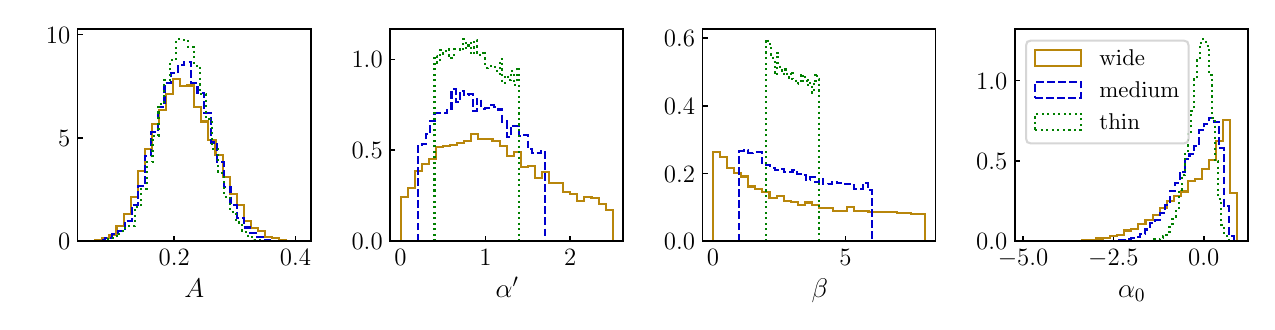}
\caption{All results and labels as in Fig.~\ref{posteriors_model} except for $d$ quarks.}
\label{posteriors_model_d}
\end{figure*}

In Fig.~\ref{posteriors_model_d}, we present the posterior distributions for the model fit of the $d$ quark results. We note that the $\alpha'$ posteriors have a less well-defined Gaussian form when compared to the $u$ quark results, Fig.~\ref{posteriors_model}. Further, the $\beta$ parameter is largely unconstrained, but in contrast to the $u$ quark, the $d$ quark results for this parameter are not as strongly skewed to the lower bound.

\begin{figure*}[t!] 
\includegraphics[width=0.92\linewidth]{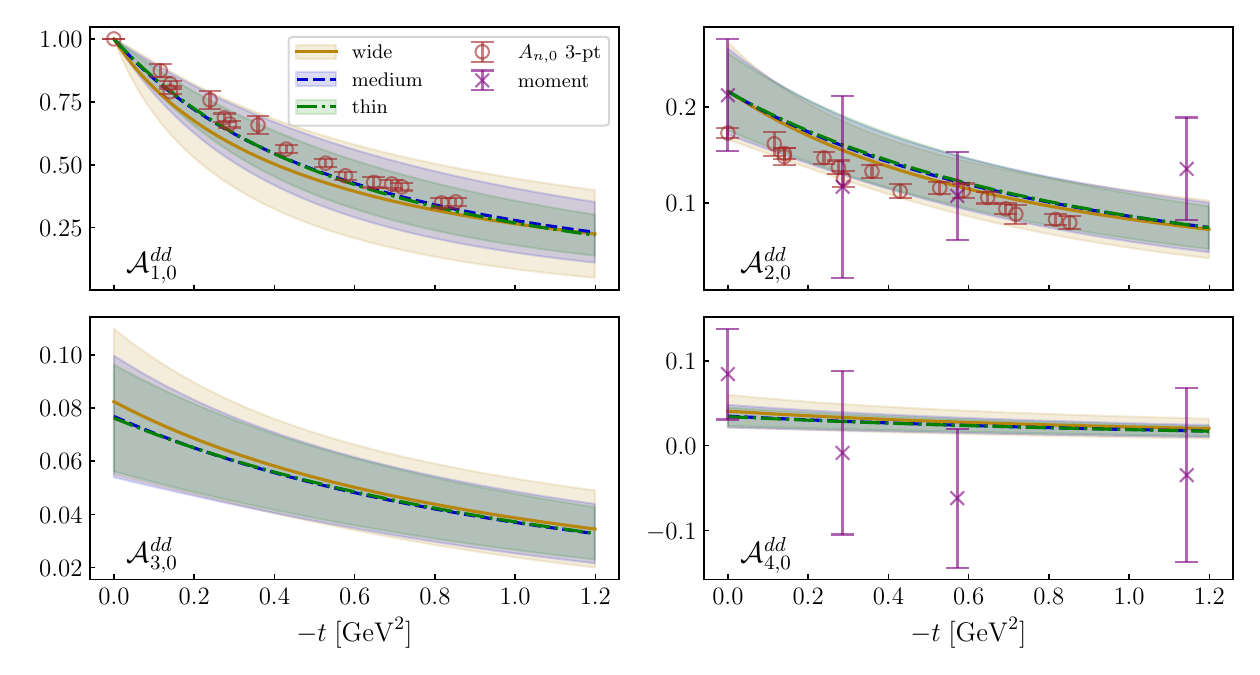}
\caption{All results and labels as in Fig.~\ref{A_u_allmoms} except for $d$ quarks.}
\label{A_d_allmoms}
\end{figure*}

In contrast to the $u$ quark results (see Fig.~\ref{w_allfits}), the $d$ quark model fits appear to have less dependence on the prior distribution---i.e.~the `wide', `medium' and `thin' results show better agreement. This can also be seen in Fig.~\ref{A_d_allmoms}, where we plot the results for the moments of the $d$ quarks. 

Similar to the $u$ quark results, we note reasonable agreement between the model fits, the direct moment fits and the leading-twist GFFs determine from the local twist-two operators (labelled `$A_{n,0}$ 3-pt'). Moreover, the $\mathcal{A}_{4,0}$ moment is not well-determined compared to $\mathcal{A}_{2,0}$.

\begin{figure*}[t!] 
\includegraphics[width=0.92\linewidth]{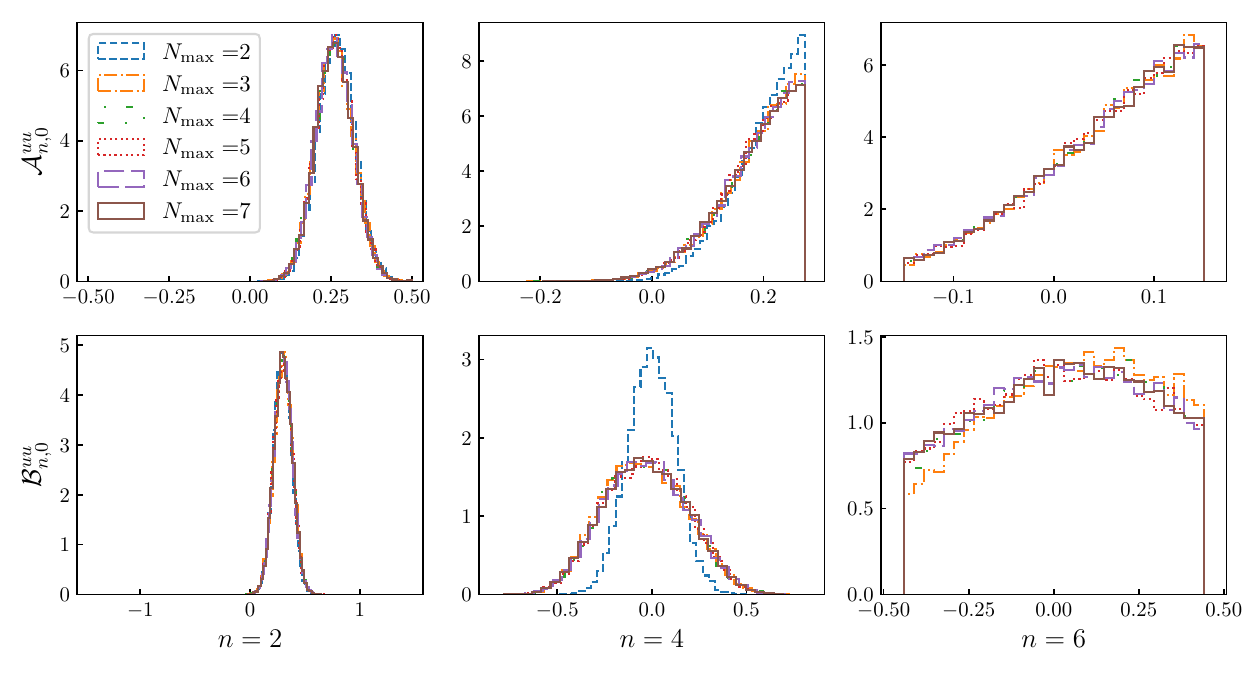}
\caption{Posterior distributions for the moments of the off-forward structure functions $\mathcal{H}_1$ (top) and $\mathcal{E}_1$ (bottom) using the fit function in Eq.~\eqref{modelindependent_fitfun}. Here $N_{\text{max}}$ is the number of moments included in the fits, with $n=2N_{\text{max}}$ as the highest moment. The range of the uniform priors is given in Eq.~\eqref{offfwdpriors}. All results for $t=-0.57\;\text{GeV}^2$ and $uu$ quark combination only.}
\label{moment_posteriors}
\end{figure*}

\section{Mellin Moment Fit} \label{sec:appendixmoments}

For the direct  fits to the Mellin moments, using the fit function Eq.~\eqref{modelindependent_fitfun}, we use uniform prior distributions given by the constraints in Eq.~\eqref{offfwdpriors}.

The posteriors of these fits are presented in Fig.~\ref{moment_posteriors} for $t=-0.57\;\text{GeV}^2$ and $u$ quarks. We observe that the order of truncation, $N_{\text{max}}$ has negligible effect on the $n=2,4$ moments for $N_{\text{max}}>2$. However, for the $\mathcal{A}_{n,0}$ moments, we note that the $n=4,6$ the distributions are highly skewed towards the upper bound. This suggests that the bounds in Eq.~\eqref{offfwdpriors} are over constraining for our results. This could be the result of (1) systematics such as the lack of large $\bomega$ results or discretisation artefacts, or (2) that the GPD positivity bound Eq.~\eqref{GPDpositivity} is broken at $\bar{Q}^2\approx 5\;\text{GeV}^2$. At our current precision and control of systematics, we can not draw strong conclusions. However, this demonstrates a possible application for our data in testing theoretical GPD constraints at moderately large $\bar{Q}^2$.

\section{Generalised form factors from local operators} \label{sec:appendix3pt}

\maketitle
\begin{figure}[t!]
    \centering
    \includegraphics[width = 0.92\textwidth]{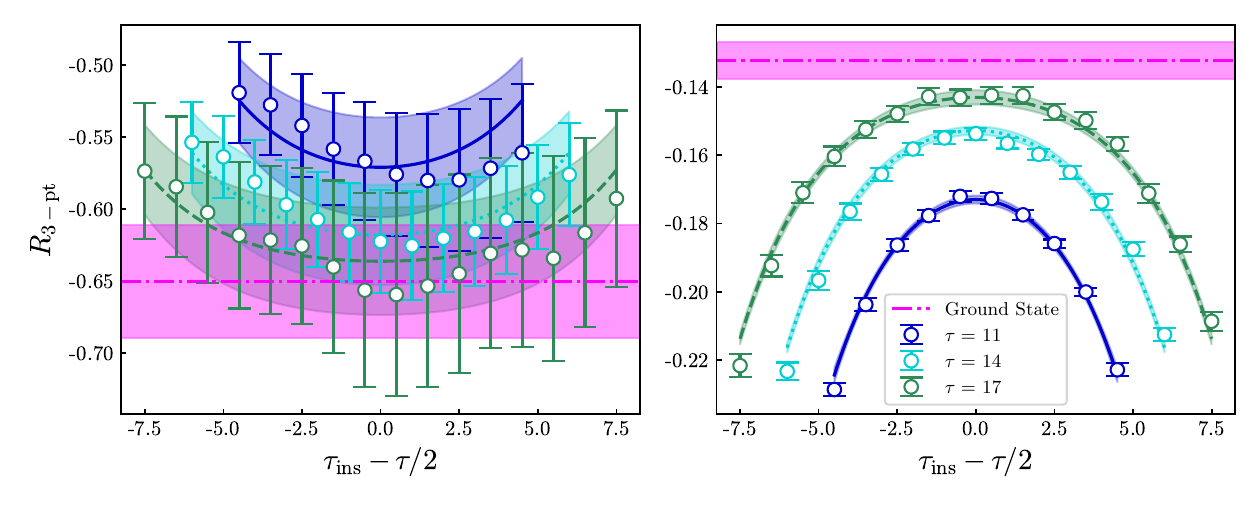}
    \caption{Euclidean time dependence of the ratio in Eq.~\eqref{eq: ratio} for the operator $
    \bar{\psi}\gamma_4\psi$ (left) and the operator $\mathcal{O}_{v_2b}$ (right). Both ratios have the unpolarised spin-parity projector and sink momenta $\mathbf{p}'=(1,0,0)$, which implies $\mathbf{\Delta}^2=0$.}
    \label{fig: ratio fits}
\end{figure}

At various points in this work, we have compared our Compton structure function moments, $\mathcal{A}_{n,0}$ and $\mathcal{B}_{n,0}$, to the leading-twist generalised form factors $A_{n,0}$ and $B_{n,0}$ (see Figs.~\ref{moments}, \ref{A_u_allmoms} and \ref{A_d_allmoms}). We determine these generalised form factors using the standard technique of calculating the matrix elements of twist-two local operators using three-point functions.

For $A_{1,0}$ and $B_{1,0}$ (note we do not compare $B_{1,0}$ to our Compton amplitude moments), we use all components of the local vector current:
\begin{align}
    j_{\mu} = \bar{\psi}\gamma_{\mu}\psi, \quad \text{for } \mu=1,2,3,4.
\end{align}
For $A_{2,0}$ and $B_{2,0}$ we use the operators
\begin{equation}
 \mathcal{O}_{v_2 a} = \frac{1}{2}\sum_{i=1}^3 \left ( \mathcal{O}_{4i}^V+\mathcal{O}_{i4}^V \right) \quad \text{and} \quad   \mathcal{O}_{v_2 b} = \mathcal{O}_{44}^V - \frac{1}{3}\sum_{i=1}^3 \mathcal{O}_{ii}^V,
\end{equation}
where
\begin{equation}
    \mathcal{O}_{\mu\nu}^V = \overline{\psi}\gamma_\mu i \overleftrightarrow{D}_\nu \psi.
\end{equation}
To determine the matrix elements of these operators, we compute the three-point correlation function:
\begin{equation}
    \mathcal{G}^{\Gamma}_{3-\text{pt}}(\mathbf{p}',\tau;\mathbf{\Delta},\tau_{\text{ins}}) = 
    \sum_{\mathbf{x}_1,\mathbf{x}_2}e^{-i\mathbf{p}'\cdot \mathbf{x}_2}e^{i\mathbf{\Delta}\cdot\mathbf{x}_1}\Gamma^{\alpha\beta} \langle \chi_\beta(\mathbf{x}_2,\tau)\mathcal{O}(\mathbf{x}_1,\tau_{\text{ins}}){\chi}^{\dagger}_\alpha(\mathbf{0},0)\rangle,
\end{equation}
The transfer 3-momentum is defined as $\mathbf{\Delta} = \mathbf{p}'-\mathbf{p}$. The local operator, $\mathcal{O}$, is inserted at time slice $\tau_{\text{ins}}$, where $\tau > \tau_{\text{ins}} > 0$. We use the spin-parity projectors given in Eq.~\eqref{spinparproj}, making use of all polarisation directions ($x,y$ and $z$).

To isolate ground state contribution we construct the ratio
\begin{equation}
\label{eq: ratio}
    {R}_{\text{3-pt}} = \frac{\mathcal{G}_{3-\text{pt}}^{\Gamma}(\mathbf{p}',\tau;\mathbf{\Delta},\tau_{\text{ins}})}{\mathcal{G}_{2-\text{pt}}(\mathbf{p}',\tau)}\left[\frac{\mathcal{G}_{2-\text{pt}}(\mathbf{p}',\tau)\mathcal{G}_{2-\text{pt}}(\mathbf{p}',\tau_{\text{ins}})\mathcal{G}_{2-\text{pt}}(\mathbf{p},\tau-\tau_{\text{ins}})}{\mathcal{G}_{2-\text{pt}}(\mathbf{p},\tau)\mathcal{G}_{2-\text{pt}}(\mathbf{p},\tau_{\text{ins}})\mathcal{G}_{2-\text{pt}}(\mathbf{p}',\tau-\tau_{\text{ins}})} \right]^{\frac{1}{2}}.
\end{equation}
In order to control excited state contamination, we make use of a two-state ansatz for the correlation functions similar to that in Ref.~\cite{ottnadexcited} for our Euclidean time fits. See Fig.~\ref{fig: ratio fits} for selected Euclidean time fits to $R_{\text{3-pt}}$.

These calculations are performed on the same set of gauge configurations as our OFCA calculation---see Tab.~\ref{tab:gauge_details}---using $N_{\text{meas}}=1074$, making their statistics comparable to our Compton amplitude results (see Tab.~\ref{tab:q_kinematics}). We calculate the three-point correlators for 16 values of the soft momentum transfer $t=-\mathbf{\Delta}^2$. The results are multiplicatively renormalised in $\text{RI}^{\prime}\text{/MOM}$.

\twocolumngrid

\end{document}